\documentclass[a4paper,10pt]{article}
\usepackage{jheppub,lipsum,ulem,hyphenat,xcolor,tcolorbox,slashed,float,bbold,bm,comment,enumitem,multicol,multirow,mathtools,sidecap,tikz}
\usepackage[T1]{fontenc}
\usepackage{changepage}
\graphicspath{ {figures/} }
\pdfoutput=1  
\allowdisplaybreaks

\usepackage{color}

\definecolor{darkcyan}{cmyk}{1,0,0,0.4}

\definecolor{darkred}{cmyk}{0,1,1,0.4}

\definecolor{lime}{HTML}{A6CE39}
\DeclareRobustCommand{\orcidicon}{\hspace{-2.1mm}
\begin{tikzpicture}
\draw[lime, fill=lime] (0,0) circle [radius=0.13] node[white] {{\fontfamily{qag}\selectfont \tiny \,ID}}; \draw[white, fill=white] (-0.0525,0.095) circle [radius=0.007]; 
\end{tikzpicture} \hspace{-3.2mm} }
\foreach \x in {A, ..., Z}{\expandafter\xdef\csname orcid\x\endcsname{\noexpand\href{https://orcid.org/\csname orcidauthor\x\endcsname} {\noexpand\orcidicon}}}

\title{Search for exotic leptons in final states with two or three leptons and fat-jets at 13 TeV LHC}

\author[a,b]{Saiyad Ashanujjaman\orcidA{}}
\author[c]{, Debajyoti Choudhury}
\author[a,b]{, Kirtiman Ghosh}
\affiliation[a]{Institute of Physics, Bhubaneswar, Sachivalaya Marg, Sainik School Post, Bhubaneswar 751005, India}
\affiliation[b]{Homi Bhabha National Institute, Training School Complex, Anushakti Nagar, Mumbai 400094, India}
\affiliation[c]{Department of Physics and Astrophysics, University of Delhi, Delhi 110007, India}
\emailAdd{saiyad.a@iopb.res.in}
\emailAdd{debchou.physics@gmail.com}
\emailAdd{kirti.gh@gmail.com}

\abstract{Exotic leptons in large gauge multiplets, appearing in many scenarios beyond the Standard Model (SM), can be produced at the LHC in pairs or association. Owing to their large masses, their eventual decay products---SM leptons and bosons---tend to be highly boosted, with the jets stemming from the SM bosons more likely to manifest themselves as a single fat-jet rather than two resolved ones. With the corresponding SM backgrounds being suppressed, final states with two or three leptons and one or two fat-jets are expected to be sensitive in probing exotic fermions much heavier than 1 TeV, and we propose and investigate an appropriate search strategy. To concentrate on the essential, we consider extensions of the SM by leptonic multiplets of a single kind (triplets, quadruplets or quintuplets), bearing in mind that such simplified models typically arise as low-energy limits of more ambitious scenarios addressing various lacunae of the SM. Performing a systematic and comprehensive study of nine such scenarios at the 13 TeV LHC, we find that the corresponding $5\sigma$ discovery reaches a range from 985 GeV to 1650 GeV (1345 GeV to 2020 GeV) for 300 (3000) fb$^{-1}$.}

\keywords{Doubly charged leptons, Exotic leptons, Fat-jet signatures.}
\preprint{IP/BBSR/2021-7}
\notoc

\begin{document} 
\maketitle
\flushbottom                                                                                                              

\section{\label{sec:intro}Introduction}
Notwithstanding the Standard Model's (SM's) impressive success in describing the fundamental interactions, it falls short of offering explanations for quite a few theoretical issues as well as some experimental observations. Fermion mass and mixing hierarchies, electroweak vacuum instability, the strong CP problem and the naturalness problem are amongst the pressing theoretical issues. Experimental inconsistencies include issues like neutrino masses and mixings, the anomalous magnetic moments of the electron and the muon, several anomalies in the $b$-quark sector, the baryon asymmetry in the Universe and the dark matter. Multifarious new physics models going beyond the SM (BSM) have been proposed to address these shortcomings. A noteworthy feature of many of these BSM models is the prediction of new particles at the TeV scale.

A wealth of BSM models addressing the issue of neutrino masses and mixings such as the canonical see-saw models and their variants \cite{Minkowski:1977sc,Yanagida:1979as,Gell-Mann:1979vob,Glashow:1979nm,Mohapatra:1979ia,Foot:1988aq,Bajc:2006ia,FileviezPerez:2007bcw,FileviezPerez:2007yji,Bonnet:2009ej,Liao:2010ku,Bonnet:2012kz,Cepedello:2017lyo,Anamiati:2018cuq,Arbelaez:2019cmj,Picek:2009is,Kumericki:2011hf,Kumericki:2012bh,McDonald:2013hsa,Avnish:2020rhx,Agarwalla:2018xpc,Li:2009mw,Ashanujjaman:2021jhi,Ashanujjaman:2020tuv,Babu:2009aq,Delgado:2011iz,Ma:2013tda,Ma:2014zda,Yu:2015pwa,Ashanujjaman:2021zrh} envisage the presence of new leptonic gauge multiplets. For instance, Ref.~\cite{Babu:2009aq} extended the fermion sector by vector-like $SU(2)_L$ triplet leptons with hypercharge $Y=1$ in order to generate
neutrino masses via effective dimension seven operators. Similarly, Refs.~\cite{Anamiati:2018cuq,Ashanujjaman:2020tuv}, in generating neutrino masses via effective dimension nine operators, introduced several fermionic multiplets, namely vector-like $SU(2)_L$ triplets with $Y=1$, vector-like $SU(2)_L$ quadruplets with $Y=1/2$ and chiral $SU(2)_L$ quintuplets with $Y=0$. The presence of several charged exotic leptons within the new weak gauge leptonic multiplets is a common yet salient feature of most of these neutrino mass models \cite{Bonnet:2009ej,Liao:2010ku,Bonnet:2012kz,Kumericki:2012bf,Cepedello:2017lyo,Anamiati:2018cuq,Arbelaez:2019cmj,Picek:2009is,Kumericki:2011hf,Kumericki:2012bh,Yu:2015pwa,McDonald:2013hsa,Delgado:2011iz,Ma:2013tda,Ma:2014zda,Ko:2015uma,Avnish:2020rhx,Agarwalla:2018xpc,KumarAgarwalla:2018nrn,Kumar:2019tat,Ashanujjaman:2020tuv,Babu:2009aq,Li:2009mw,Ashanujjaman:2021jhi,Kumar:2021umc}. Copious production of different such charged exotic leptons, driven by their electroweak couplings to the $W$ and $Z$ bosons and the photon, and their eventual decays to the SM leptons and bosons ($W/Z/h$-boson) offer interesting ways to probe them at the Large Hadron Collider (LHC). Phenomenology of such exotics leptons (in particular, the doubly-charged one) at the LHC has been studied extensively in the literature \cite{Alloul:2013raa,Delgado:2011iz,Ma:2013tda,Chen:2013xpa,Ding:2014nga,Ma:2014zda,Yu:2015pwa,Leonardi:2014epa,Biondini:2012ny,Kumar:2021umc}. However, all of these searches considered multileptons, missing transverse momentum ($p_T^{\rm miss}$) and jets in the final states. Typically, these final state signatures are beset with considerably large SM background contributions. Moreover, for signatures with $p_T^{\rm miss}$, multiple leptons and jets with unidentifiable origins, kinematic reconstructions of the exotic leptons. Consequently, these searches are often non-optimal in probing exotic leptons with masses beyond 1 TeV. This behoves us to inspect complementary signatures. For TeV scale exotics, their decay products (SM leptons and bosons) could be sufficiently boosted that the jets emanating from them would be collimated. Consequently, the hadronically decaying bosons are more likely to manifest as a single fat-jet than two resolved jets. On the contrary, only a small fraction of the SM background events would give rise to fat-jets in the final state. Not only are signatures comprising multileptons and fat-jets cleaner than those in the searches mentioned above \cite{Alloul:2013raa,Delgado:2011iz,Ma:2013tda,Chen:2013xpa,Ding:2014nga,Ma:2014zda,Yu:2015pwa,Leonardi:2014epa,Biondini:2012ny}, but most of these final states also allow for kinematic reconstructions of the exotic leptons. Consequently, such signatures are expected to be sensitive in probing exotic leptons with masses beyond 1 TeV.

Adopting the bottom-up approach for new physics, we consider several simplified models by extending the SM with only one type of new leptonic weak gauge multiplet at a time. We design a search for the charged exotic leptons belonging to these simplified models in final states with multilepton and fat-jets. By performing a systematic and comprehensive collider study, we estimate the $5\sigma$ discovery reach of this search for nine such simplified models for both 300 and 3000 fb$^{-1}$ of the integrated luminosity of proton-proton collision data at the 13 TeV LHC.

The rest of this work is structured as follows. In Sec.~\ref{sec:models}, we briefly discuss some simplified models. Productions and decays of the exotic leptons are discussed in Sec.~\ref{sec:prod_dec}. We perform a systematic and comprehensive collider study in Sec.~\ref{sec:collider} followed by a summary in Sec.~\ref{sec:conclusion}.

\section{\label{sec:models}Simplified models}
Understandably, most recent BSM searches by both the CMS and ATLAS collaborations have been designed to probe specific models that are well-placed to address particular shortcomings of the SM. Being tuned to a specific model, such strategies often lose sensitivity when applied to more general models but with similar final states. This behoves us to design a search strategy for charged exotic leptons resulting in final states with multilepton and fat-jets so that the former could be applied to any BSM scenario that includes such exotic leptons decaying into each other or an SM lepton and a boson without impinging much on the sensitivity. To this end, we consider several simplified models\footnote{Such simplified models typically arise as low-energy limits of more ambitious scenarios \cite{Anamiati:2018cuq,Arbelaez:2019cmj,Picek:2009is,Kumericki:2011hf,Kumericki:2012bh,Yu:2015pwa,McDonald:2013hsa,Delgado:2011iz,Ma:2013tda,Ma:2014zda,Avnish:2020rhx,Agarwalla:2018xpc,Ashanujjaman:2020tuv,Babu:2009aq} addressing various shortcomings of the SM. The simplified models, thus, involve a smaller set of parameters compared to the more complete ones.} adopting the bottom-up approach for new physics.

Keeping the SM gauge group unaltered, we supplement the SM field content with exotic leptonic states. While a fully general analysis is virtually impossible, we attempt a generic framework by allowing the new leptonic states to belong to different $SU(2)_L$ representations, namely {\bf 3}, {\bf 4} and {\bf 5}. As for the hypercharge assignments, we limit these to (1,2), (1/2,3/2,5/2) and (0,1,2,3) for the triplet, quadruplet and quintuplet fields, respectively. This restriction is imposed to ensure that these multiplets have at least one component of both doubly- and singly-charged fields. Therefore, we have nine different leptonic multiplets, namely $3^F_1$, $3^F_2$, $4^F_{1/2}$, $4^F_{3/2}$, $4^F_{5/2}$, $5^F_0$, $5^F_1$, $5^F_2$ and $5^F_3$, and these, along with their differently charged components, are listed in Table~\ref{table:model}. In our notation, $\chi_Y^Q$ stands for the $Q$-charged component of the $SU(2)_L$ $\chi$-plet with hypercharge $Y$.

\begin{table}[h]
\centering
\begin{tabular}{l}
\hline
\\
$3^F_1={\begin{pmatrix*}[l] 3_1^{++} \\ 3_1^{+} \\ 3_1^0 \end{pmatrix*}}$, \quad
$3^F_2={\begin{pmatrix*}[l] 3_2^{3+} \\ 3_2^{++} \\ 3_2^+ \end{pmatrix*}}$, \quad
$4^F_{1/2}={\begin{pmatrix*}[l] 4_{1/2}^{++} \\ 4_{1/2}^{+} \\ 4_{1/2}^0 \\ 4_{1/2}^- \end{pmatrix*}}$, \quad
$4^F_{3/2}={\begin{pmatrix*}[l] 4_{3/2}^{3+} \\ 4_{3/2}^{++} \\ 4_{3/2}^+ \\ 4_{3/2}^0 \end{pmatrix*}}$, \quad
$4^F_{5/2}={\begin{pmatrix*}[l] 4_{5/2}^{4+} \\ 4_{5/2}^{3+} \\ 4_{5/2}^{++} \\ 4_{5/2}^+ \end{pmatrix*}}$, \quad
\\
\\
$5^F_{0}={\begin{pmatrix*}[l] 5_0^{++} \\ 5_0^{+} \\ 5_0^{0} \\ 5_0^- \\ 5_0^{--} \end{pmatrix*}}$, \quad
$5^F_{1}={\begin{pmatrix*}[l] 5_1^{3+} \\ 5_1^{++} \\ 5_1^{+} \\ 5_1^0 \\ 5_1^{-} \end{pmatrix*}}$, \quad
$5^F_{2}={\begin{pmatrix*}[l] 5_2^{4+} \\ 5_2^{3+} \\ 5_2^{++} \\ 5_2^+ \\ 5_2^{0} \end{pmatrix*}}$, \quad
$5^F_{3}={\begin{pmatrix*}[l] 5_3^{5+} \\ 5_3^{4+} \\ 5_3^{3+} \\ 5_3^{++} \\ 5_3^{+} \end{pmatrix*}}$.
\\
\\ 
\hline
\end{tabular}
\caption{\label{table:model} Exotic leptonic multiplets and their components.}
\end{table}

With the LHC having already constrained such leptons to be heavier than a few hundred GeV, clearly, such masses cannot be generated through electroweak symmetry breaking, as this would not only entail uncomfortably large Yukawa couplings but also generate too large a diphoton branching fraction for the SM Higgs. Thus, bare mass terms need to be present, and multiplets with non-zero hypercharges must be vector-like, allowing for a Dirac mass term $M_{\chi_Y} \overline{\chi^F_{YL}} \chi^F_{YR}$. This would
automatically cancel the $U(1)$ anomaly without introducing additional states. Multiplets with vanishing hypercharges (and, hence, no contribution to the anomaly) could very well be chiral, yet allowing for a lepton number violating mass term such as $M_{\chi_Y} \overline{\left(\chi^F_{YR}\right)^c} \chi^F_{YR}$. Here onwards, we omit the subscript $Y$ from the mass term for brevity. Being gauge invariant, these masses (whether lepton-number conserving or violating) would naturally be comparable to the new physics scale. The differently charged components in a given multiplet are mass-degenerate at the tree-level. Radiative corrections, dominantly induced by the
electroweak gauge bosons, lift this degeneracy and induce mass-splittings between the differently charged components $\chi^Q_Y$ and $\chi^{Q^\prime}_Y$ \cite{Cirelli:2005uq}
\begin{equation*}
\Delta M (\chi^Q_Y,\chi^{Q^\prime}_Y) \approx \frac{\alpha M_\chi}{4\pi} \left[(Q^2-Q^{\prime 2})  f\left(r_Z\right) + \frac{Q-Q^\prime}{\sin^2 \theta_w}(Q+Q^\prime-2Y)\left\{ f\left(r_W\right)-f\left(r_Z\right) \right\} \right]~,
\end{equation*}
where $\theta_w$ is the weak-mixing angle, $\alpha$ is the electromagnetic fine structure constant, $r_a \equiv m_a/m_\chi$, and the function $f(r)$ has the form
\begin{equation*}
f(r) =\frac{r}{2} \left[2r^3 \ln r
-2r+\sqrt{r^2-4}~(r^2+2) \ln \left\{\frac{r^2-2-r\sqrt{r^2-4}}{2} \right\} \right]~.
\end{equation*}
For the mass range of our interest ($M_\chi \gg m_{W,Z}$), the function $f(r) \approx - 2 \pi r$ with the imaginary part being suppressed by more than an order of magnitude. Consequently, the mass splittings being completely determined by the electric- and hyper-charges of the particles involved are almost independent of the common mass. Most importantly, the components with larger charges are the heavier ones. Summarised in Table~\ref{table:mass-splitting}, these splittings\footnote{It might be argued that the tree-level mass splittings introduced by possible Higgs couplings would be larger. Given that we would be including exotic multiplets of only one type, such couplings are relevant only for the $3^F_1$ field. Even there, concerns such as neutrino masses and possible FCNCs restrict these couplings to be tiny and the splittings generated thereby to be of little consequence.} are inconsequential as far as production cross-sections at the LHC are concerned. However, these play a crucial role in their decays; we defer this discussion till Sec.~\ref{sec:prod_dec}.

\begin{table}[h]
\centering
\scalebox{0.7}{
\begin{tabular}{l}
\hline
\\
\begin{minipage}{.38\textwidth}
$\Delta M (3_1^{++},3_1^+) \sim$ [860--870] MeV,
\end{minipage}%
\begin{minipage}{.38\textwidth}
$\Delta M (3_1^{+},3_1^0) \sim$ [515--525] MeV;
\end{minipage}%
\\
\\
\begin{minipage}{.38\textwidth}
$\Delta M (3_2^{3+},3_2^{++}) \sim$ [1545--1570] MeV,
\end{minipage}%
\begin{minipage}{.38\textwidth}
$\Delta M (3_2^{++},3_2^+) \sim$ [1200--1220] MeV;
\end{minipage}%
\\ 
\\
\begin{minipage}{.38\textwidth}
$\Delta M (4_{1/2}^{++},4_{1/2}^{+}) \sim$ [685--695] MeV,
\end{minipage}%
\begin{minipage}{.38\textwidth}
$\Delta M (4_{1/2}^{+},4_{1/2}^{0}) \sim$ [340--350] MeV,
\end{minipage}%
\begin{minipage}{.38\textwidth}
$\Delta M (4_{1/2}^{0},4_{1/2}^{-}) \sim$ [-5--2] MeV;
\end{minipage}%
\\ 
\\
\begin{minipage}{.38\textwidth}
$\Delta M (4_{3/2}^{3+},4_{3/2}^{++}) \sim$ [1370--1390] MeV,
\end{minipage}%
\begin{minipage}{.38\textwidth}
$\Delta M (4_{3/2}^{++},4_{3/2}^{+}) \sim$ [1025--1045] MeV,
\end{minipage}%
\begin{minipage}{.38\textwidth}
$\Delta M (4_{3/2}^{+},4_{3/2}^{0}) \sim$ [685--700] MeV;
\end{minipage}%
\\
\\
\begin{minipage}{.38\textwidth}
$\Delta M (4_{5/2}^{4+},4_{5/2}^{3+}) \sim$ [2060--2090] MeV,
\end{minipage}%
\begin{minipage}{.38\textwidth}
$\Delta M (4_{5/2}^{3+},4_{5/2}^{++}) \sim$ [1715--1745] MeV,
\end{minipage}%
\begin{minipage}{.38\textwidth}
$\Delta M (4_{5/2}^{++},4_{5/2}^{+}) \sim$ [1370--1400] MeV;
\end{minipage}%
\\
\\
\begin{minipage}{.38\textwidth}
$\Delta M (5_0^{\pm \pm},5_0^{\pm}) \sim$ [515--520] MeV,
\end{minipage}%
\begin{minipage}{.38\textwidth}
$\Delta M (5_0^{\pm},5_0^{0}) \sim$ 170 MeV;
\end{minipage}%
\\
\\
\begin{minipage}{.38\textwidth}
$\Delta M (5_1^{3+},5_1^{++}) \sim$ [1200--1220] MeV,
\end{minipage}%
\begin{minipage}{.38\textwidth}
$\Delta M (5_1^{++},5_1^{+}) \sim$ [855--870] MeV,
\end{minipage}%
\begin{minipage}{.38\textwidth}
$\Delta M (5_1^{+},5_1^{0}) \sim$ [515--525] MeV,
\end{minipage}%
\begin{minipage}{.35\textwidth}
$\Delta M (5_1^{0},5_1^{-}) \sim$ 170 MeV;
\end{minipage}%
\\
\\
\begin{minipage}{.38\textwidth}
$\Delta M (5_2^{4+},5_2^{3+}) \sim$ [1890--1915] MeV,
\end{minipage}%
\begin{minipage}{.38\textwidth}
$\Delta M (5_2^{3+},5_2^{++}) \sim$ [1545--1570] MeV,
\end{minipage}%
\begin{minipage}{.38\textwidth}
$\Delta M (5_2^{++},5_2^{+}) \sim$ [1200--1220] MeV,
\end{minipage}%
\begin{minipage}{.35\textwidth}
$\Delta M (5_2^{+},5_2^{0}) \sim$ [855--875] MeV;
\end{minipage}%
\\
\\
\begin{minipage}{.38\textwidth}
$\Delta M (5_3^{5+},5_3^{4+}) \sim$ [2580--2615] MeV;
\end{minipage}%
\begin{minipage}{.38\textwidth}
$\Delta M (5_3^{4+},5_3^{3+}) \sim$ [2230--2265] MeV,
\end{minipage}%
\begin{minipage}{.38\textwidth}
$\Delta M (5_3^{3+},5_3^{++}) \sim$ [1885--1920] MeV,
\end{minipage}%
\begin{minipage}{.35\textwidth}
$\Delta M (5_3^{++},5_3^{+}) \sim$ [1540--1570] MeV;
\end{minipage}%
\\
\\
\hline
\end{tabular}
}
\caption{\label{table:mass-splitting} Mass-splittings between the differently charged components of the exotics multiplets.}
\end{table}

In this work, we extend the SM with a single copy of only one of the aforementioned multiplets at a time. In other words, we consider nine simple scenarios, each of which we denote as {\it Model-$\chi^F_Y$} depending on the identity of the exotic multiplets (see Table \ref{table:model}).  Such simplified models can often be viewed as low-energy limits of specific well-motivated BSM models when some of the fields decouple and/or some of the interactions cease to exist (see Table \ref{table:refofmodels}). For instance, {\it Model-}$3^F_1$ is a limit of the neutrino mass model presented in Ref.~\cite{Babu:2009aq} when the scalar quadruplet with $Y=3/2$ decouples. Likewise, {\it Model-}$3^F_1$ ({\it Model-}$4^F_{1/2}$) [{\it Model-}$5^F_0$] could be thought of as a limit of another neutrino mass model in Ref.~\cite{Ashanujjaman:2020tuv} when the $4^F_{1/2}$ and $5^F_{0}$ ($3^F_1$ and $5^F_{0}$) [$3^F_1$ and $4^F_{1/2}$] multiplets get decoupled. Understandably though, such a correspondence is not unique. Moreover, a given scenario, say {\it Model-}$5^F_3$, may not be easily relatable to any popular theory.

\begin{table}[htb!]
\centering
\scalebox{1.0}{
\begin{tabular}{|c|c|c|}
\hline 
\multirow{2}{*}{Simplified model} & \multicolumn{2}{|c|}{Corresponding more ambitious scenario}\\
\cline{2-3}
& Refs. & Motivation \\
\hline \hline
\multirow{2}{*}{{\it Model-}$3^F_{1}$} & \cite{Babu:2009aq} & neutrino mass \\
& \cite{Delgado:2011iz,Ma:2013tda,Ashanujjaman:2020tuv} & neutrino mass \\
\hline
\multirow{2}{*}{{\it Model-}$4^F_{1/2}$} & \cite{McDonald:2013hsa} & neutrino mass \\
& \cite{Ashanujjaman:2020tuv} & neutrino mass \\
\hline
\multirow{3}{*}{{\it Model-}$5^F_0$} & \cite{Cai:2011qr,Kumar:2021umc} & neutrino mass and dark matter \\
& \cite{Kumericki:2012bf,Kumericki:2012bh,Yu:2015pwa,Kumar:2019tat,Ashanujjaman:2020tuv} & neutrino mass \\
& \cite{Cirelli:2009uv,Ko:2015uma,KumarAgarwalla:2018nrn} & dark matter \\
\hline
\multirow{2}{*}{{\it Model-}$5^F_1$} & \cite{Picek:2009is,Kumericki:2011hf} & neutrino mass \\
& \cite{KumarAgarwalla:2018nrn,Ko:2015uma} & dark matter \\
\hline
{\it Model-}$5^F_2$ & \cite{KumarAgarwalla:2018nrn,Ko:2015uma} & dark matter \\
\hline \hline
\end{tabular} 
}
\caption{\label{table:refofmodels}Some simplified models and corresponding more ambitious scenarios studied in the literature.} 
\end{table}

Irrespective of such correspondence, each of these simplified models is UV-complete in its own right. And while some of the exotics can play a direct role in generating neutrino masses \cite{Yu:2015pwa,McDonald:2013hsa,Delgado:2011iz,Ma:2013tda,Ma:2014zda,Avnish:2020rhx,Agarwalla:2018xpc,Picek:2009is,Kumericki:2011hf,Kumericki:2012bh,Babu:2009aq,Anamiati:2018cuq,Arbelaez:2019cmj,Ashanujjaman:2020tuv} and/or addressing other concerns like the anomalous magnetic moments of the muon and the electron or even the possible evidence of lepton flavour violation in $B$-decays, the others may not play a role in facilitating these. In particular, in order to generate three non-zero masses for the light neutrinos, most of the neutrino mass models \cite{Bonnet:2009ej,Liao:2010ku,Bonnet:2012kz,Cepedello:2017lyo,Anamiati:2018cuq,Arbelaez:2019cmj,Picek:2009is,Kumericki:2011hf,Kumericki:2012bh,Yu:2015pwa,McDonald:2013hsa,Delgado:2011iz,Ma:2013tda,Ma:2014zda,Avnish:2020rhx,Agarwalla:2018xpc,Ashanujjaman:2020tuv,Babu:2009aq,Li:2009mw,Ashanujjaman:2021jhi} require three copies of the exotic multiplets. While it might be argued that, in all strictness, as of now, only two of the light neutrino states need to be massive, and thus only two of the heavy generations are called for, it should be realised that were more sensitive experiments to conclude that all the light neutrino species are indeed massive, one would need three exotic species. However, most extant experimental searches \cite{ATLAS:2018ghc,CMS:2019lwf,ATLAS:2020wop,ATLAS:2021xxb,CMS:2021zkl,ATLAS:2021eyc} consider one such species, and thus an easy comparison with such studies is facilitated by the inclusion of just one species rather than two or three.

All the simplified models considered in this work accommodate large $SU(2)_L$ multiplet. The introduction of such multiplet changes the running of the $U(1)_Y$ and $SU(2)_L$ gauge couplings ($g_1$ and $g_2$). In Table~\ref{table:rge} (in Appendix~\ref{app:A}), we list the one-loop $\beta$-functions describing the evolution of $g_1$ and $g_2$ in these models. Also tabulated are the energy scales at which the gauge couplings blow up (the Landau poles in $g_1$ and/or $g_2$), scales that vary strongly across these models. The said energy scale is well below the grand unification scale for some of these models. For instance, the Landau pole in $g_2$ appear at a scale of $10^{10}$ GeV for all the vector-like quintuplet models. For the {\it Model-$5^F_3$}, the Landau pole in $g_1$ appears at a scale of $10^6$ GeV. In other words, these specific models need to be superseded well below such scales.

\section{\label{sec:prod_dec}Production and decays of the exotics}
The exotic leptons are pair produced aplenty at the LHC by quark-antiquark annihilation through $s$-channel $\gamma/Z$ and $W^\pm$ exchanges:
\[
q\bar{q^\prime} \to W^* \to \chi^{Q} \chi^{-Q\pm 1} \quad {\rm and} \quad q\bar{q} \to \gamma^*/Z^* \to \chi^Q \chi^{-Q}.
\]
They can also be pair produced via the $t/u$-channel photon-photon fusion process,\footnote{They can also be pair produced via vector boson-fusion ($\gamma Z, \gamma W^\pm, ZZ, ZW^\pm$) processes, with one or two associated forward jets at the LHC. However, such processes are rather sub-dominant, and can be safely neglected.} namely
\[
\gamma \gamma \to \chi^Q \chi^{-Q} \quad (Q \neq 0).
\]
Despite their very small (as compared to the other partons) density, photon initiated pair production becomes significant for large masses of the multi-charged exotics \cite{Babu:2016rcr,Ghosh:2017jbw}, with the bulk of the contribution emanating from subprocesses wherein one of the photons originates from a resolved proton. And while single production (in association with an SM lepton) is, in principle, possible, such processes are highly suppressed on account of the small mixing.

We evaluate the leading order (LO) production cross-sections of the exotics at the 13 TeV LHC using the UFO modules generated from \texttt{SARAH} \cite{Staub:2013tta,Staub:2015kfa} in \texttt{MadGraph} \cite{Alwall:2011uj,Alwall:2014hca} with the {\it LUXqed17-plus-PDF4LHC15-nnlo-100} parton distribution function (PDF) \cite{Manohar:2016nzj,Manohar:2017eqh,Butterworth:2015oua}.\footnote{The {\it LUXqed17} PDF \cite{Manohar:2016nzj,Manohar:2017eqh}, the photon PDF inside the proton determined in a model-independent manner using electron-proton scattering data, combined with QCD partons from the {\it PDF4LHC15} PDF \cite{Butterworth:2015oua} results in the {\it LUXqed17-plus-PDF4LHC15-nnlo-100} PDF.}'\footnote{In order to estimate `true' LO production cross-section, one should, in principle, convolute the LO partonic cross-section with the LO partonic distributions only \cite{Basu:2002uu}, and thus the extensively used {\it NNPDF23-LO-as-0130-qed} PDF should have been an obvious choice. However, as pointed out in Refs.~\cite{Ghosh:2018drw,Fuks:2019clu}, in the case of this PDF, the predicted production cross-sections for the photon fusion initiated processes suffer from substantial uncertainties due to large uncertainties in the photon density inside the proton. On the contrary, the {\it LUXqed17-plus-PDF4LHC15-nnlo-100} PDF has much smaller errors in determining the photon density inside the proton. Moreover, for the mass range of our interest, both the PDFs predicts almost the same Drell-Yan production cross-sections---the difference being well within a few per cent. In view of these, we chose the latter PDF over the former one.}

\begin{figure}[htb!]
\centering
\includegraphics[width=0.45\columnwidth]{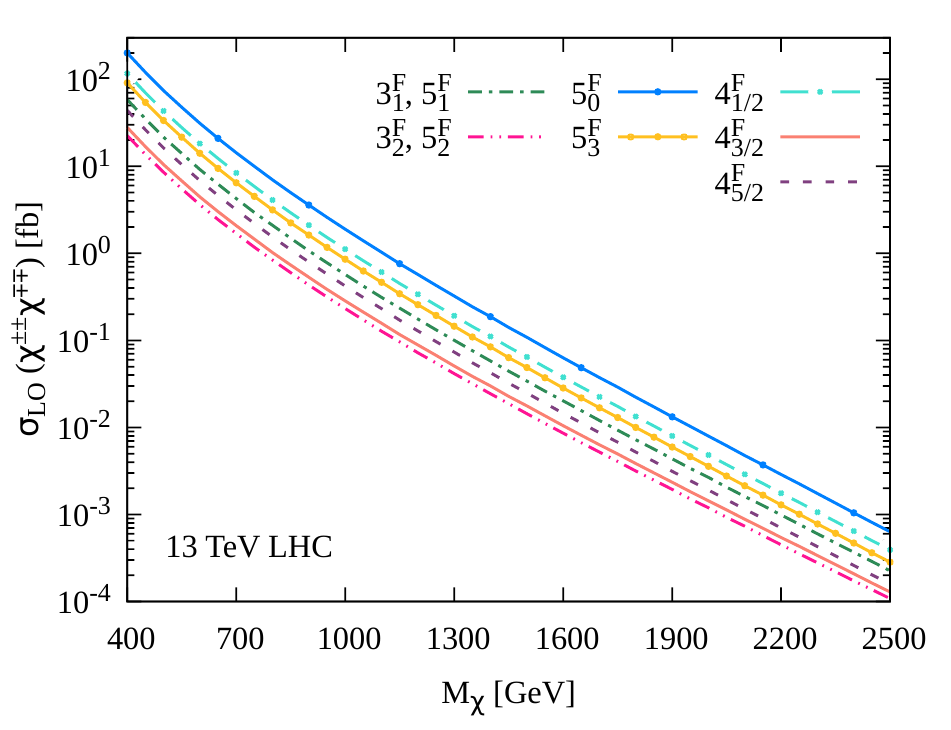} \quad
\includegraphics[width=0.45\columnwidth]{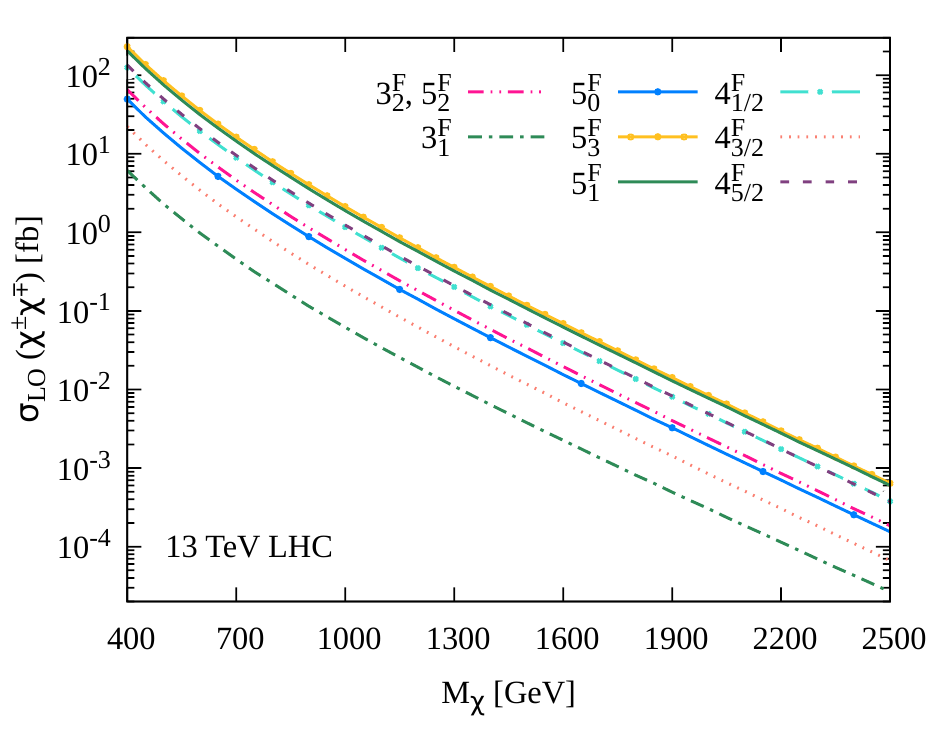}
\includegraphics[width=0.45\columnwidth]{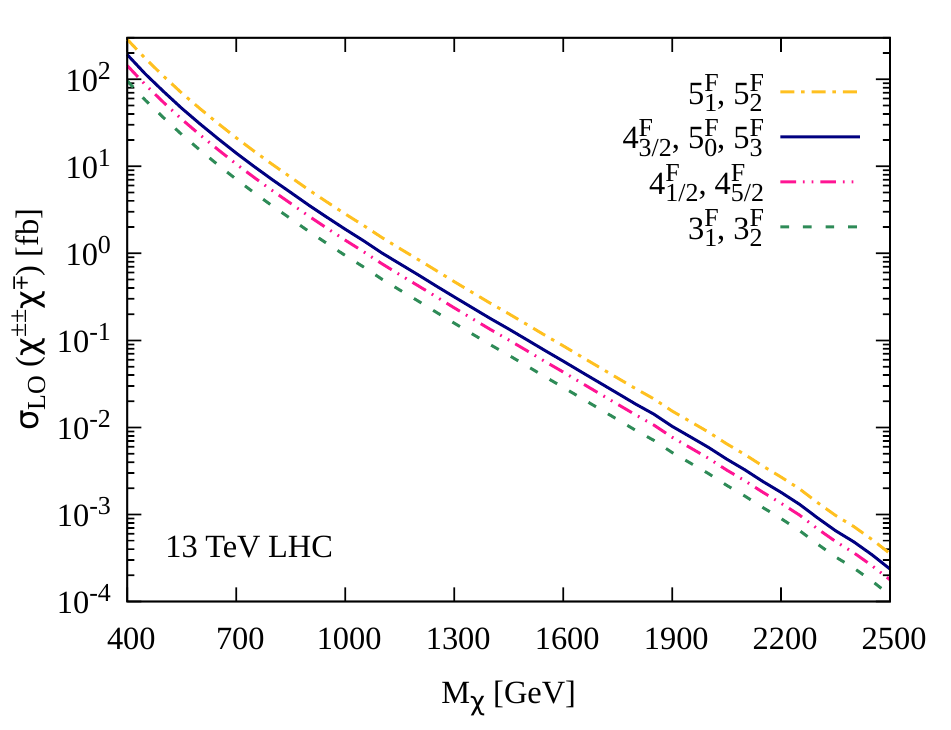} \quad
\includegraphics[width=0.45\columnwidth]{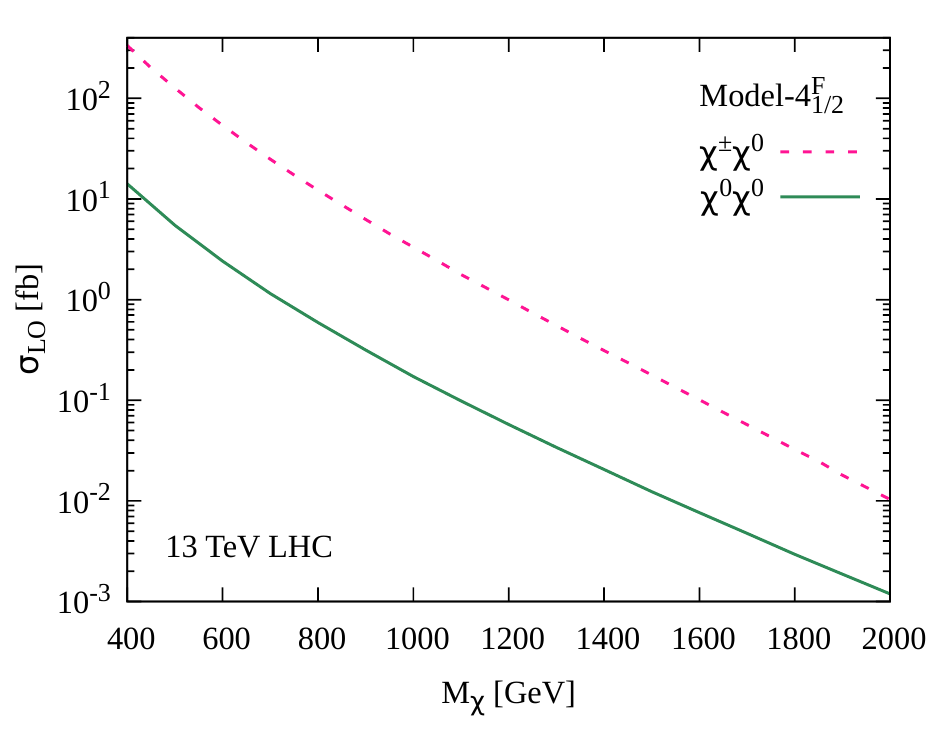}
\caption{LO production cross-sections of the exotics in different simplified models: doubly-charged pair production (top left), singly-charged pair production (top right), doubly- and singly-charged associated production (bottom left), and neutral pair and neutral-charged productions (bottom right).}
\label{fig:cs}
\end{figure}

The plots in Fig.~\ref{fig:cs} show different pair and associated LO production cross-sections for the exotic leptons at the 13 TeV LHC as a function of their tree-level
mass for the different scenarios considered in this work. The left (right) plot in the top panel shows the pair production cross-sections for the doubly- (singly-) charged exotics, while the left plot in the bottom panel shows the associated production cross-sections for the doubly- and singly-charged exotics.\footnote{In models with higher-charged leptons as well, those too would be produced, and we return to this issue later.}'\footnote{Some of the simplified models considered in this work also contain neutral leptons (see Table~\ref{table:model}). However, we do not consider them, partly because their pair production cross-sections are smaller than those for the charged ones (see the right plot in the bottom panel of Fig.~\ref{fig:cs}). Though in some of these models, their productions in association with the singly-charged ones are comparable to those of the other processes, with signal regions in the present search being designed for the charged exotic leptons, the inclusion of the neutral ones would not entail significant added contributions. For instance, trilepton events from charged exotics production would demonstrate a peak in the invariant mass distribution of a particular lepton---distinguishable from the others---and the fat-jet, whereas this is not the case for those from neutral production. In view of these and the existing search paradigm (that the charged ones are likely to be discovered prior to the neutral ones) adopted by the ATLAS/CMS collaborations, we only consider the charged exotic leptons.}Refs.~\cite{Fuks:2012qx,Fuks:2013vua,Ruiz:2015zca} have estimated the QCD corrections to the production of heavy $SU(2)_L$ triplet leptons at hadron colliders. The resulting next-to-leading order (NLO) $K$-factors varies from 1.1 to 1.4 for both charged current and neutral current processes over a triplet mass range of 100 GeV to 2 TeV. Further, Ref.~\cite{Ruiz:2015zca} shows that the NLO differential $K$-factors for heavy lepton kinematics are substantially flat, suggesting that naive scaling by an overall $K$-factor is a very good approximation. And while the corrections have not been calculated for the higher multiplets, it is obvious that the said corrections pertain to the hadronic ends of the Feynman diagrams, and these are universal for all the cases under discussion. Guided by this, we apply an overall QCD $K$-factor of 1.25 to the LO cross-section.

The decays of the exotics often depend on the details of the UV-completion, for example, through their effective mixings with the SM counterparts, either through direct
Yukawa couplings or through a combination of couplings involving yet other exotics. In broad terms, the leading decays can be categorised into three classes:
\begin{itemize}
\item {\it Category-I:} The aforementioned mixing can induce two-body decays into a SM lepton and a boson. The rates are, understandably, dependent on the elusive parameters. 
\item {\it Category-II:} The radiative mass-splittings between the differently charged components within a multiplet induces the heavier exotics to decay into the lighter ones in association with $\ell^+ \nu$, $\pi^+$, $K^+$, $\pi^+ \pi^0/\rho(770)$, {\it etc.} if kinematically accessible; we refer to these decays as heavy state transitions. Driven by gauge interactions, these are governed only by mass splittings.
\item {\it Category-III:} It could go to into a $W$-boson and an off-shell (exotic) partner which goes into an SM lepton and an SM boson. Thus, in effect, we have a 3-body decay into two bosons and an SM lepton.
\end{itemize}

The dominance of one category of decays over the others depends, naturally, on the parameter space.  For example, being suppressed by both the mixing and the exotic lepton mass in the propagator, the rates for {\it Category-III} are, typically, much smaller \cite{Kumericki:2011hf} than the {\it Category-II} decays and can be safely neglected. Similarly, for the singly- and doubly-charged exotic leptons, the corresponding mass-splittings are not large enough ($\sim$ a few hundred MeV -- 1 GeV) so that the {\it Category-I} decays dominate over the {\it Category-II} ones. On the contrary, for the multi-charged (with $|Q| \geq 3$) exotic leptons, {\it Category-I} decays are not allowed at all. Consequently, these undergo {\it Category-II} decays. In summary, the dominant decay modes for the exotic leptons are
\begin{equation*}
\chi^{\pm \pm} \to \ell^\pm W^\pm \quad {\rm and} \quad
\chi^{\pm} \to \ell^\pm Z, \, \ell^\pm h, \, \nu W^\pm \quad {\rm with} \quad \ell = e,\mu,\tau~,
\end{equation*}
while for $n \geq 3$,
\[
\chi^{n\pm} \to \chi^{(n-1)\pm} + \hat X \, \quad {\rm with} \quad \hat X \in \pi^\pm, \, K^\pm, \, \pi^\pm \pi^0, \, \rho^\pm(770), \, \ell^\pm \nu.
\]

Although the relevant mass-splittings (see Table~\ref{table:mass-splitting}) are small, these are substantial enough to ensure that some of the heavy state transitions are prompt. However, the hadrons/leptons stemming from such decays are very soft, and thus can not be easily identified (as the resultant of a primary decay) in the LHC environment owing to the inherent large hadronic activity. Therefore, multi-charged exotics' pair and associated productions serve only to enhance the doubly-charged pair production effectively.  For instance, the effective pair production cross-section of doubly-charged exotics in {\it Model-$3^F_2$} is given by
\begin{equation*}
\sigma^{\rm eff} (3^{\pm \pm}_2 3^{\mp \mp}_2) = \sigma(3^{\pm \pm}_2 3^{\mp \mp}_2) + \sigma(3^{3\pm}_2 3^{3\mp}_2) + \sigma(3^{3\pm}_2 3^{\mp \mp}_2)~.
\end{equation*}

\begin{figure}[htb!]
\centering
\includegraphics[width=0.45\columnwidth]{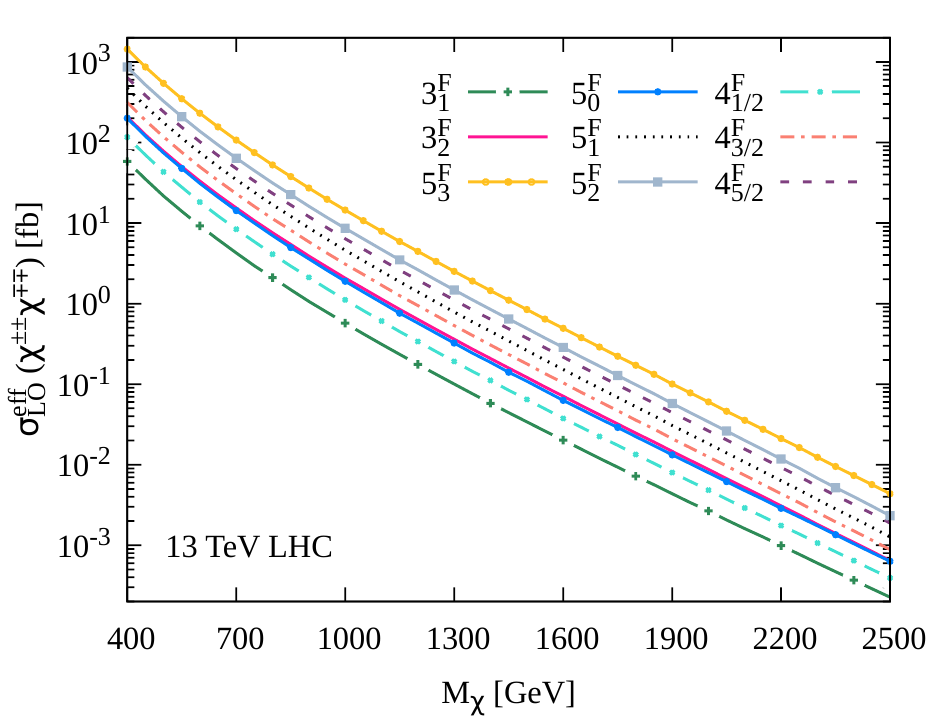}
\caption{Effective cross-section of doubly-charged pair production at LO in different simplified models.}
\label{fig:effcs}
\end{figure}

We plot the LO effective cross-sections for doubly-charged pair production as a function of their tree-level mass in different simplified models in Fig.~\ref{fig:effcs}.  We see that all the models, particularly those with multiplet lying in the higher representation of $SU(2)_L$, yield sizeable cross-sections at the 13 TeV LHC.

For simplicity, we assume equal branching fractions\footnote{This assumption does not have a substantial impact on our final results. For one, both the $W$ and $Z$ boson have similar hadronic branching fractions; second, the analysis to be presented here does not distinguish between the fat-jets emanating from them.} for the decays $\chi^\pm \to \ell^\pm Z$ and $\chi^\pm \to \ell^\pm h$. Further, to facilitate an easy comparison with existing search paradigms adopted by the ATLAS/CMS collaborations \cite{ATLAS:2018ghc,CMS:2019lwf,ATLAS:2020wop,ATLAS:2021xxb}, we assume that their branching fractions are identical across all lepton flavours ({\it flavour-democratic scenario}).

\section{\label{sec:collider}Collider phenomenology}
The exotic leptons are copiously produced via neutral or charged current Drell-Yan-like processes at the LHC. As mentioned earlier, the multi-charged exotic leptons decay promptly into the doubly-charged exotic partner in association with mesons/leptons, which are very soft, and thus are not likely to pass the reconstruction criteria of the LHC detectors. Therefore, the production of the multi-charged exotics effectively enhances the cross-sections for the doubly-charged ones. Subsequent prompt decays to SM particles lead to various final state signatures at the LHC. Listed in Table~\ref{table:signatures}, these include smoking gun signatures like mono-Higgs, di-Higgs, di-gauge boson, {\it etc.} in association with multiple leptons. Further decays of the SM bosons lead to various final states, some of which allow for kinematic reconstructions of the exotics. The exotics are expected to be in the TeV regime, so their decay products (SM leptons and bosons) would be highly boosted. Consequently, the jets emanating from the decaying SM bosons would tend to be collimated, with such a boson more likely to manifest itself as a single fat-jet (with a larger radius) rather than two resolved jets. Though many of the final states are interesting in their own right, we emphasise only those with multilepton and fat-jets.

\begin{table}[htb!]
\centering
\scalebox{1.0}{
\begin{tabular}{|c||c|c|c|c|} 
\hline
& $\bm{\chi^{++} \to \ell^{'+} W^+}$ & $\bm{\chi^{+} \to \nu W^+}$ & $\bm{\chi^{+} \to \ell^{'+} Z}$ & $\bm{\chi^{+} \to \ell^{'+} h}$ \\
\hline
\hline
$\bm{\chi^{--} \to \ell^- W^-}$ & $\ell^- \ell^{'+} W^- W^+$ & $\ell^- \nu W^- W^+$ & $\ell^- \ell^{'+} W^- Z$ & $\ell^- \ell^{'+} W^- h$ \\
\hline
$\bm{\chi^{-} \to \nu W^-}$ & $\nu \ell^{'+} W^- W^+ $ & $\nu \nu W^- W^+ $ & $\nu \ell^{'+} W^- Z $ & $\nu \ell^{'+} W^- h $ \\
\hline
$\bm{\chi^{-}  \to \ell^- Z}$ & $\ell^- \ell^{'+} Z W^+$ & $\ell^- \nu Z W^+$ & $\ell^- \ell^{'+} Z Z$ & $\ell^- \ell^{'+} Z h$ \\
\hline
$\bm{\chi^{-} \to \ell^- h}$ & $\ell^- \ell^{'+} h W^+$ & $\ell^- \nu h W^+ $ & $\ell^- \ell^{'+} h Z $ & $\ell^- \ell^{'+} h h$ \\
\hline
\end{tabular}
}
\caption{Possible final state signatures of pair/associated productions of charged exotic leptons.}
\label{table:signatures}
\end{table}

In the following, we briefly review the processes contributing to the final state signatures of our interest.

\begin{itemize}
\item Trilepton and fat-jet: Both pair and associated production of doubly-charged exotics contribute to this final state as follows:
\begin{enumerate}[label=(\alph*)]
\item $\chi^{++} \chi^{--} \to \ell^+ W^+ \ell^{'-} W^- \to \ell^+ \ell^{'-} \ell^{''\pm} \nu J$,
\item $\chi^{\pm \pm} \chi^\mp \to \ell^\pm W^\pm \ell^{'\mp} Z/h \to \ell^\pm \ell^{'\pm} \ell^{''\mp} \nu J$,
\end{enumerate}
where $J$ denotes a fat-jet emanating from the hadronically decaying SM boson. Here, (one of) the $W$-boson(s) from the $\chi^{\pm \pm}$ decays leptonically and the other $W(Z/h)$ from the $\chi^{\pm \pm}(\chi^\pm)$ decays hadronically. Hence, two same-charge ({\it sc}) leptons result from the $\chi^{\pm \pm}$ decay, and the other exotic
fermion results in an opposite-charge ({\it oc}) lepton and a hadronically decaying boosted boson. Thus, not only would the mass of the fat-jet peak at $M_{W/Z/h}$, but the mass of $\chi^{\pm \pm}$ can also be reconstructed from the peak of the invariant mass distribution of the opposite-charge lepton and the fat-jet.

\item Same-charge dilepton and fat-jet: Only associated production of doubly-charged exotics contribute to this final state:
\begin{enumerate}[label=(\alph*)]
\item[{}] $\chi^{\pm \pm} \chi^\mp \to \ell^\pm W^\pm \nu W^\mp \to \ell^\pm \ell^{' \pm} \nu \nu J$.
\end{enumerate}

In this case, the $W$-boson from the $\chi^{\pm \pm}$ decays leptonically while that from the $\chi^\pm$ decays hadronically. Since neither of $\chi^{\pm \pm}$ nor $\chi^\pm$ decays to entirely visible final states, and since there are two neutrinos in the final state, none of two masses can be reconstructed.

\item Opposite-charge dilepton and fat-jets: All possible production of doubly- and singly-charged exotics contribute to this final state\footnote{Note that the opposite-charge dilepton final states usually have huge SM backgrounds. However, the signal rates are also large because of contributions from all possible production
channels. Hoping to reduce the background contributions to a manageable level using appropriate kinematic cuts, we consider these final states.} as follows:
\begin{enumerate}[label=(\alph*)]
\item $\chi^{\pm \pm} \chi^\mp \to \ell^\pm W^\pm \nu W^\mp \to \ell^\pm \ell^{'\mp} \nu \nu J$,
\item $\chi^+ \chi^- \to \ell^\pm Z/h \; \nu W^\mp \to \ell^\pm \ell^{'\mp} \nu \nu J$,
\item $\chi^{++} \chi^{--} \to \ell^+ W^+ \ell^{'-} W^- \to \ell^{'+} \ell^{'-} J J$,
\item $\chi^{\pm \pm} \chi^\mp \to \ell^\pm W^\pm \ell^{'\mp} Z/h \to \ell^\pm \ell^{'\mp} J J$,
\item $\chi^+ \chi^- \to \ell^+ Z/h \ell^{'-} Z/h \to \ell^+ \ell^{'-} J J$.
\end{enumerate}

For processes (a) and (b), the $W$($Z/h$)-boson from the $\chi^{\pm \pm}$($\chi^\pm$) decays hadronically, and the other $W$ from the $\chi^+$ decays leptonically. The lepton from the decay of $\chi^{\pm \pm}/\chi^\pm$ usually has a larger $p_T$ than that coming from the $W$-decay. Therefore, the peak of the invariant mass distribution of the leading lepton and the fat-jet system is expected to yield $M_{\chi^{\pm \pm}}$ (or $M_{\chi^\pm}$, as the case may be). For processes (c), (d) and (e), both the $W/Z/h$-bosons from $\chi^{\pm \pm}/\chi^\pm$ decay hadronically so that the final state contains two fat-jets. For such events, the reconstruction of both the exotic fermions is possible, although there are two independent pairs of invariant masses: $(i)~ M_{J_0 \ell_0}$ and $M_{J_1 \ell_1}$, and $(ii)~M_{J_1 \ell_0}$ and $M_{J_0 \ell_1}$, where $\ell_0$ and $\ell_1$ ($J_0$ and $J_1$) denote, respectively, the leading and subleading leptons (fat-jets). Since the mass difference between dissimilar exotics is expected to be small (see Table \ref{table:mass-splitting}), we may safely assume the pairing with the smaller mass difference to be the correct one.
\end{itemize}

In what follows, we briefly describe the reconstruction and selection of various physics objects, event selection, and the classification of selected events into mutually exclusive signal regions (SRs).

\subsection{\label{sec:object}Object reconstruction and selection}
We use \texttt{Delphes} \cite{deFavereau:2013fsa} to reconstruct different physics objects, namely photons, electrons, muons and jets. For reasons to be elaborated later (in Sec.~\ref{sec:SR}), for each event, we attempt jet reconstruction with two different values for the distance parameter ($R$), namely $0.8$ (applicable for fat-jets) and $0.4$ (for ordinary jets). The jet constituents are clustered using the {\it anti-k$_T$ algorithm} \cite{Cacciari:2008gp} as implemented in \texttt{FastJet} \cite{Cacciari:2011ma}, with a winner-take-all axis being used for the fat-jets. At the reconstruction level, fat-jets and ordinary-jets (denoted by $J$ and $j$, respectively) are not dissociated objects; rather a pair of ordinary-jets may coalesce into a single fat-jet on increasing $R$, and, conversely, a single fat-jet may resolve into two ordinary jets if a small $R$ is used.\footnote{Obviously, both kinds of jets are not used for a given event.} The {\it jet pruning} algorithm \cite{Ellis:2009su,Ellis:2009me} is used to remove the soft yet wide-angle QCD emissions from the fat-jets,. Following Ref.~\cite{Ellis:2009su,Ellis:2009me}, we choose the default values for the parameters of the pruning algorithm, namely $z_{cut} = 0.1$ and $R_{cut} = 0.5$.\footnote{The parameters $z_{cut}$ and $R_{cut}$ determine the level of pruning---the larger (smaller) the $z_{cut}$ ($R_{cut}$), the more aggressive is the pruning. Pruning with a smaller $R_{cut}$ (larger $z_{cut}$) degrades the putative jet mass resolution by aggressively pruning the QCD shower of the daughter partons. For a larger $R_{cut}$ (smaller $z_{cut}$), the algorithm merges uncorrelated soft radiations from the underlying event and pile-up with the putative jet, thereby augmenting the jet mass. Ref.~\cite{Ellis:2009su,Ellis:2009me} demonstrated that the choice $z_{cut} = 0.1$ and $R_{cut} = 0.5$ yields close to optimal results of pruning, and the result is largely insensitive to small changes in the values of the cuts.} Further, {\it $N$-subjettiness}\footnote{An inclusive jet shape variable, N-subjettiness is a good measure of the number of putative subjets a jet resolves into. It is defined as $\tau_N = \frac{1}{d_0}\sum_k p_{T,k} {\rm min} \left\{\Delta R_{1,k},\Delta R_{2,k},...,\Delta R_{N,k} \right\}$, where $N$ is the number of subjets a jet is presumably composed of, $k$ runs over the constituent particles in a given jet $j$ with $p_{T,k}$ being their transverse momenta, and $\Delta R_{j,k}$ the distance in the rapidity-azimuth plane between a candidate subjet $j$ and a constituent particle $k$. Here, $d_0 = \sum_k p_{T,k} R_0$ is the normalisation factor with $R_0$ being the characteristic jet radius used in the original jet clustering algorithm. The ratio $\tau_N/\tau_{N-1}$ is an useful discriminant between events with $N$- and $(N-1)$-prong jets \cite{Thaler:2010tr,Thaler:2011gf}.} \cite{Thaler:2010tr,Thaler:2011gf} is used to divulge the multi-prong nature of the fat-jets. We choose {\it one-pass $k_T$ axes} for the minimal axes with a thrust parameter $\beta=1$. Reconstructed jets are required to be central (pseudorapidity range $|\eta|<2.5$) and have a transverse momentum $p_T > 30$ GeV. Lepton candidates (electrons and muons) are considered for further analysis only if these satisfy both $p_T > 10$ GeV and $|\eta|<2.5$. Lepton isolation is ensured by demanding that the scalar sum of the $p_T$s of photons and hadrons within a cone of $\Delta R=0.5$ around the lepton be smaller than 12\%(15\%) of the electron (muon) $p_T$. By truncating the hadronic activity inside the isolation cone, this requirement subdues the reducible backgrounds such as $Z+$jets and $tt+$jets, where additional leptons emanate from heavy quark decays or jet misidentification. Finally, the missing transverse momentum vector $\vec p_T^{\rm ~miss}$ (with magnitude $p_T^{\rm miss}$) is estimated using all reconstructed particle-flow objects in an event.

To circumvent possible ambiguities in assigning particles to jets, we require the latter to be separated by $\Delta R(j,j) > 0.4$. Since the jet reconstruction algorithm may also reconstruct leptons as jets, any jet within a cone of $\Delta R<0.4$ of a selected lepton is discarded to avoid double counting. Similarly, all selected electrons within a cone of $\Delta R<0.05$ of a selected muon are thrown away as these could arise due to the muon bremsstrahlung interactions with the inner detector material. To mitigate the effect of jet momenta mismeasurement contributing to missing transverse momenta, the leading fat-jet $J_0$ (or the leading and sub-leading jets $j_0$ and $j_1$) should satisfy $|\Delta \phi(J_0,p_T^{\rm miss})| > 0.2$ ($|\Delta \phi (j_{0,1},p_T^{\rm miss})| > 0.2$). It needs to be realised that some of the jets, escorted by loosely isolated leptons or hadrons that reach the muon detectors traversing the hadronic calorimeter or hadronic showers that deposit a significant fraction of energy in the electromagnetic calorimeter, could be misidentified as leptons. Though the composition of the misidentified lepton contribution varies appreciably among the analysis channels, without going into the intricacy of modelling the misidentified lepton contributions, we adopt a conservative approach, assuming the probability for a jet to be misidentified as a lepton to be in the 0.1 -- 0.3\% range \cite{ATLAS:2016iqc}.\footnote{As delineated in Ref.~\cite{ATLAS:2016iqc}, which we follow, this probability varies linearly from 0.3\% to 0.1\% for $p_T$ ranging from 20 to 50 GeV, saturating at 0.1\% for $p_T > 50$ GeV.}

Furthermore, bremsstrahlung interactions of an electron with the inner detector material could lead to deflections and kinks in its tracks which, in turn, lead to incorrect measurement of the track curvature and thus to a charge misidentification. Similarly, bremsstrahlung photons with a high enough energy may subsequently convert to $e^-e^+$ pairs within the detector material and thus give rise to additional tracks near the primary electron trajectory leading to the ambiguous association of an electromagnetic calorimeter cluster to a track. We adopt the charge misidentification probability from Ref.~\cite{ATLAS:2017xqs}: $P(p_T,\eta)=\sigma(p_T) \times f(\eta)$, where $\sigma(p_T)$ and $f(\eta)$ range, respectively, in the intervals $[0.02,0.1]$ and $[0.03,1]$ such that $P(p_T,\eta)$ ranges from 0.02\% to 10\%.\footnote{$\sigma(p_T)$ is found to vary linearly from 0.02 to 0.1 for $p_T$ ranging from 40 to 300 GeV saturating at 0.02 (0.1) for $p_T < 40$ ($p_T>300$) GeV. Similarly, $f(\eta)$ varies from 0.03 to 1.0 for $0.5 < \eta < 2.5$, while $f(\eta) \approx 0.03$ for $\eta < 0.5$.}Note that higher-energy electrons are more likely to suffer charge-misidentification owing to the smaller curvature of their tracks. Similarly, electrons coming out with a larger $|\eta|$ have a larger charge misidentification probability as they traverse through a higher amount of inner detector material and, thus, suffer more bremsstrahlung interactions.

\subsection{\label{sec:SR} Event preselection and signal regions selection}
After object reconstruction and selection, only events with one of same-sign dileptons (\textbf{\textit{SSD}}) or opposite-sign dileptons (\textbf{\textit{OSD}}) or trileptons (\textbf{\textit{3L}}) are considered for further analysis, which proceeds in two steps---the preselection and the signal region (SR) selection. Henceforth, we restrict our discussion to only those objects (leptons, jets, {\it etc.}) that pass the reconstruction and selection criteria described in Sec.~\ref{sec:object}. The preselection requirements are as follows. Events are selected only if the absolute value of the sum of charges of the energetic and isolated leptons is two, zero and one for the {\it SSD, OSD} and {\it 3L} events, respectively. To subdue background contributions from low-$\Delta R$ final-state radiations as well as low-mass resonances---$\gamma^{(*)}$ and quarkonia (such as $J/\psi, \Upsilon$)---events containing a lepton pair with $\Delta R(\ell,\ell)<0.4$ or an opposite sign same-flavour lepton pair with invariant mass below 10 GeV are discarded. In addition, events containing an opposite sign same-flavour lepton pair with an invariant mass within the nominal Z-boson mass window, {\it i.e.} $M_Z \pm 10$ GeV are vetoed. This not only suppresses background contributions from processes like $Z\to \ell \ell$ but also from processes like $Z\to \ell \ell + \gamma^{(*)} \to \ell \ell \ell^\prime \ell^\prime$ where (one or) two of the leptons fall(s) outside the detector coverage or (is) are too soft to pass the object reconstruction and selection criteria or get(s) misidentified by the detector. Furthermore, the $Z$-boson invariant mass veto is also applied to the same-sign dielectron events to subjugate the background contribution of electron charge misidentification.\footnote{This is of little concern for muons, owing to the much superior charge identification.}

Events satisfying the above-mentioned preselection criteria (denoted, henceforth, by \textbf{\textit{S0}}) are categorised into the following primary SRs ---
\begin{enumerate}[label=(\roman*)]
\item \textbf{\textit{3L-1J:}} {\it 3L} events with at least one fat-jet,
\item \textbf{\textit{SSD-1J:}} {\it SSD} events with at least one fat-jet, 
\item \textbf{\textit{OSD-1J:}} {\it OSD} events with at least one fat-jet and high-$p_T^{\rm miss} ( > 100$ GeV),
\item \textbf{\textit{OSD-2J:}} {\it OSD} events with at least two fat-jets.\footnote{At this stage, the {\it OSD-1J} and {\it OSD-2J} SRs are not mutually exclusive. However, they become mutually exclusive after all the selection cuts (see Table~\ref{table:cuts}) are imposed.}
\end{enumerate}

\noindent As we will see in Sec.~\ref{sec:eventSel}, not all the events in the above-mentioned SRs would pass the SR specific selection criteria (Table~\ref{table:cuts}), in particular, the twin criteria for a fat-jet to be identified as a ${W/Z/h}$-fat-jet (denoted as $J_{W/Z/h}$), and defined by the requirements dubbed as {\it Selection I-2} in Sec.~\ref{sec:eventSel}.\footnote{Typically, only 35\%--50\% of the fat-jets would satisfy the $J_{W/Z/h}$ selection criteria. Firstly, often some of the constituents of a putative fat-jet may go missing in the jet clustering algorithm. Second, other jets or hadrons not stemming from $W/Z/h$ could superfluously be clustered with a putative fat-jet. Further, the jet momenta mismeasurement makes the state of affairs worse.} Therefore, to supplement the sensitivity of this search, in addition to the above-mentioned SRs, we consider the following two secondary SRs corresponding to {\it 3L-1J} and {\it SSD-1J} SRs:

\begin{enumerate}[label=(\roman*)]   
\setcounter{enumi}{4}
\item \textbf{\textit{3L-2j:}} {\it 3L} events with at least two ordinary-jets, and 
\item \textbf{\textit{SSD-2j:}} {\it SSD} events with at least two ordinary-jets.
\end{enumerate}

The only difference between the last two SRs ((v) and (vi)) and their fat-jet counterparts ((i) and (ii)) is that two or more ordinary-jets are demanded instead of at least one fat-jet. In order to make these two SRs orthogonal to their fat-jet counterparts, only the events which fail to pass all the selection cuts for the {\it 3L-1J} and {\it SSD-1J} SRs are considered in the {\it 3L-2j} and {\it SSD-2j} SRs, respectively. The sole purpose of defining these two secondary SRs is, by and large, to supplement the sensitivity of this search. Note that we do not define any ordinary-jet counterpart for the {\it OSD-1J} ({\it OSD-2J}) SRs, as for such ordinary-jet SRs, the relevant backgrounds (signals) are comparatively large (small), and, thus, they are of little consequence.

\subsection{SM Backgrounds}
Having defined all the SRs of our interest, we now briefly mention the relevant backgrounds. These include diboson ($VV$ with $V=W,Z/\gamma^*$), triboson ($VVV$) and tetraboson ($VVVV$) production, Higgsstrahlung processes ($Zh$, $Wh$, $t\bar{t}h$, $ZZh$, $WWh$), multi-top production ($t\bar{t}$, $t\bar{t}t$, $t\bar{t}t\bar{t}$), top-pair in association with gauge bosons ($t\bar{t}V$, $t\bar{t}VV$), single top production ($tb,tW,tj$), and Drell-Yan processes ($W,Z/\gamma^*$). In order to meticulously account for the jet multiplicity in the final states, all the background samples are generated in association with up to two jets using \texttt{MadGraph} \cite{Alwall:2011uj,Alwall:2014hca} at the LO using the {\it five flavour scheme} followed by {\it MLM matching} in \texttt{PYTHIA} \cite{Sjostrand:2014zea}. The background event samples are normalised to NLO or higher \cite{Campbell:1999ah,Ciccolini:2003jy,Brein:2003wg,Catani:2007vq,Campanario:2008yg,Balossini:2009sa,Bredenstein:2009aj,Catani:2009sm,Kidonakis:2010ux,Campbell:2011bn,Brein:2011vx,Bevilacqua:2012em,Garzelli:2012bn,Brein:2012ne,Altenkamp:2012sx,Nhung:2013jta,Kidonakis:2013zqa,Denner:2014cla,Harlander:2014wda,Kidonakis:2015nna,Muselli:2015kba,Shen:2015cwj,Frederix:2017wme}.

The relevant backgrounds can be classified into two classes---reducible and irreducible. The leptons stemming from the SM bosons' decays, commonly referred to as prompt leptons, are usually indiscernible in momentum and isolation from those produced in signal events. Hence, all SM processes giving rise to two or more isolated leptons, such as $VV$, $VVV$, $VVVV$, $t\bar{t}$, $t\bar{t}V$ and Higgsstrahlung processes, constitute the irreducible backgrounds in this analysis. On the contrary, the reducible backgrounds---mainly from the SM processes like $Z/\gamma^*+$jets and $t\bar{t}+$jets, where a jet is misidentified as lepton or additional leptons originates from heavy quark decays---are considerably subdued by applying stringent lepton isolation requirements.

\subsection{\label{sec:eventSel} SR-specific event selection}
In this section, we discuss the SR-specific selection criteria that would be effective in subjugating the relevant SM backgrounds. We use the distributions of various kinematic variables as guiding a premise to choose the selection cuts to ameliorate the signal-to-background ratio.

\subsubsection{The {\it 3L-1J} signal}
\begin{figure}[htb!]
\centering
\includegraphics[width=0.32\columnwidth]{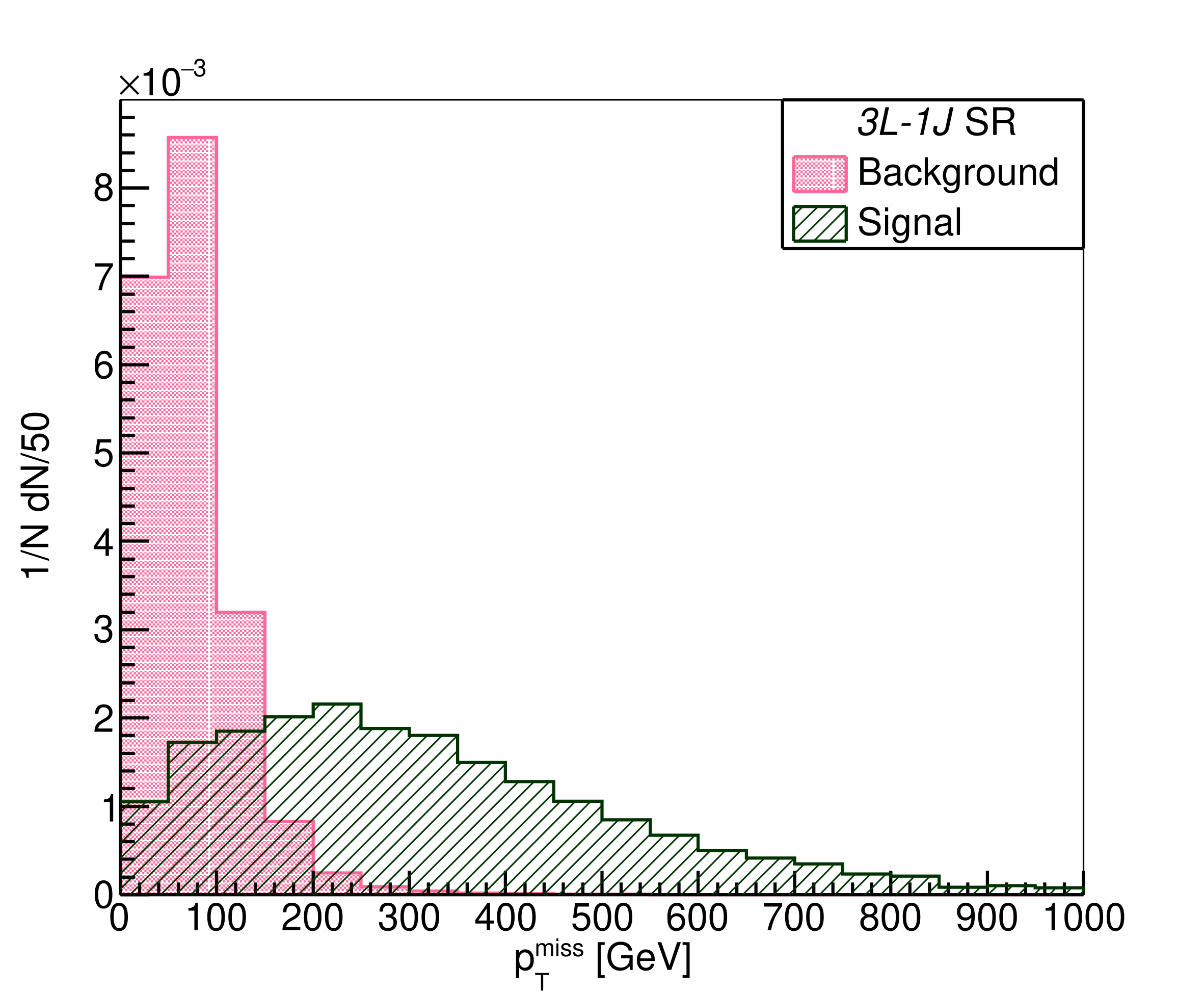}
\includegraphics[width=0.32\columnwidth]{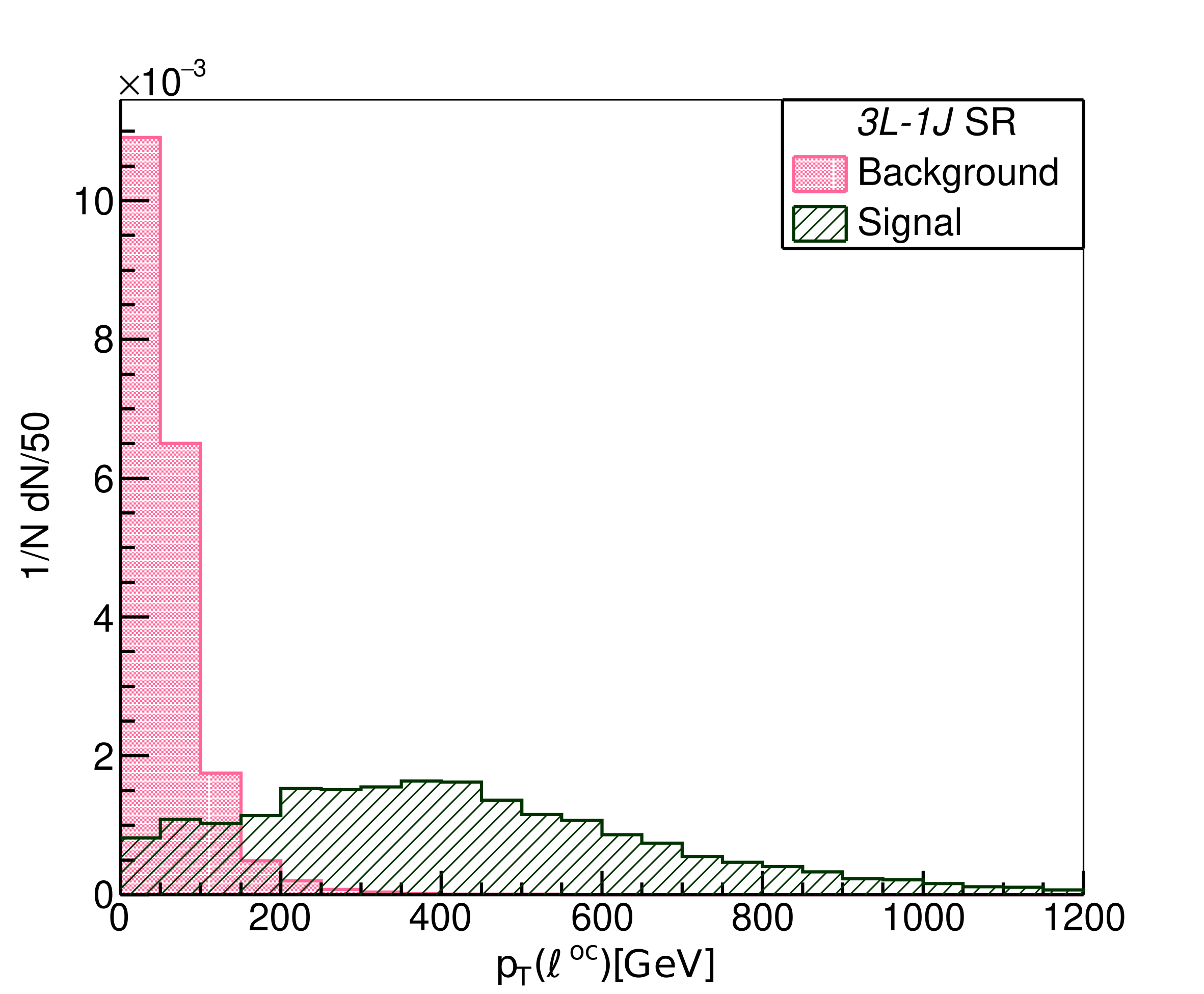}
\includegraphics[width=0.32\columnwidth]{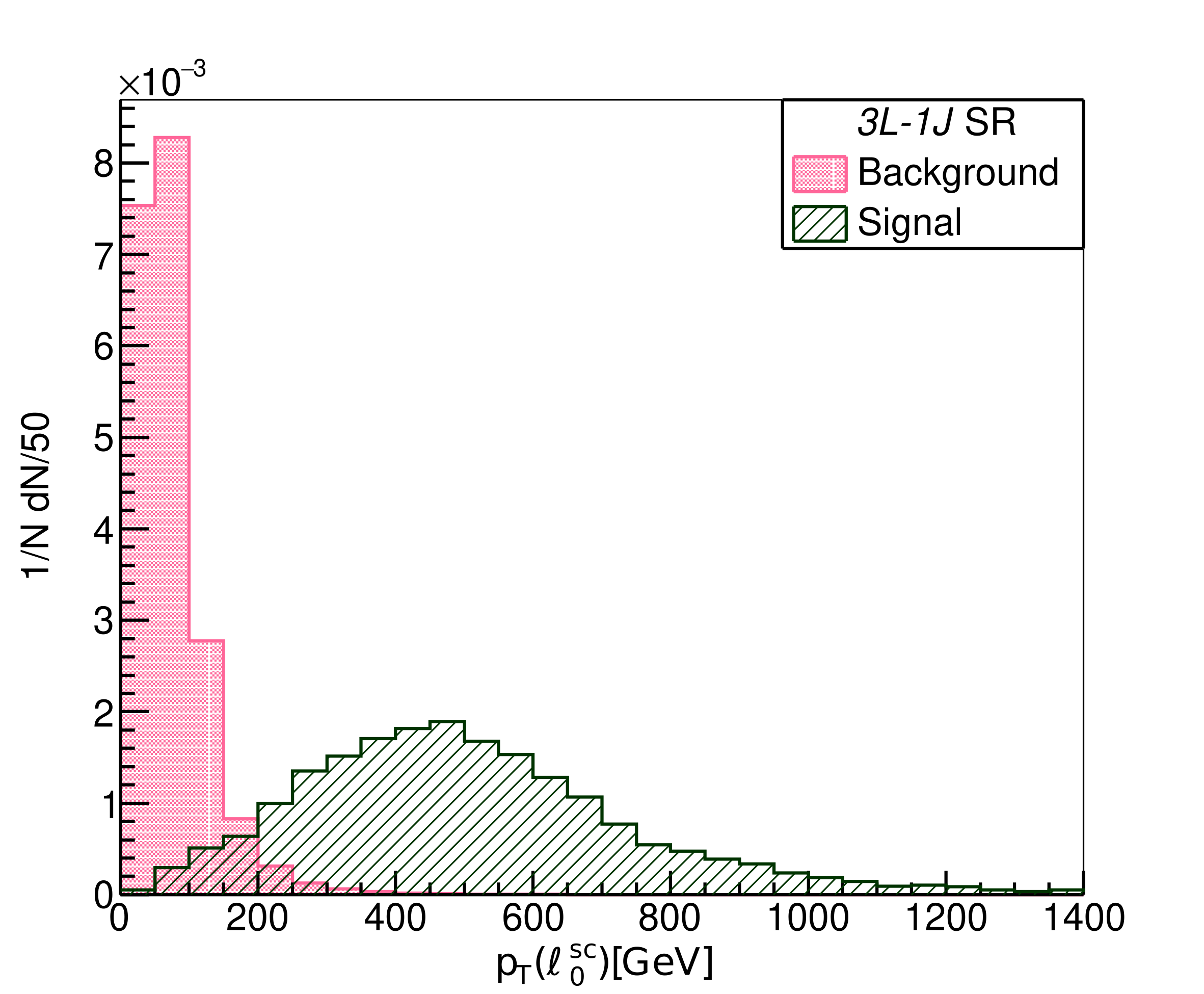}
\includegraphics[width=0.32\columnwidth]{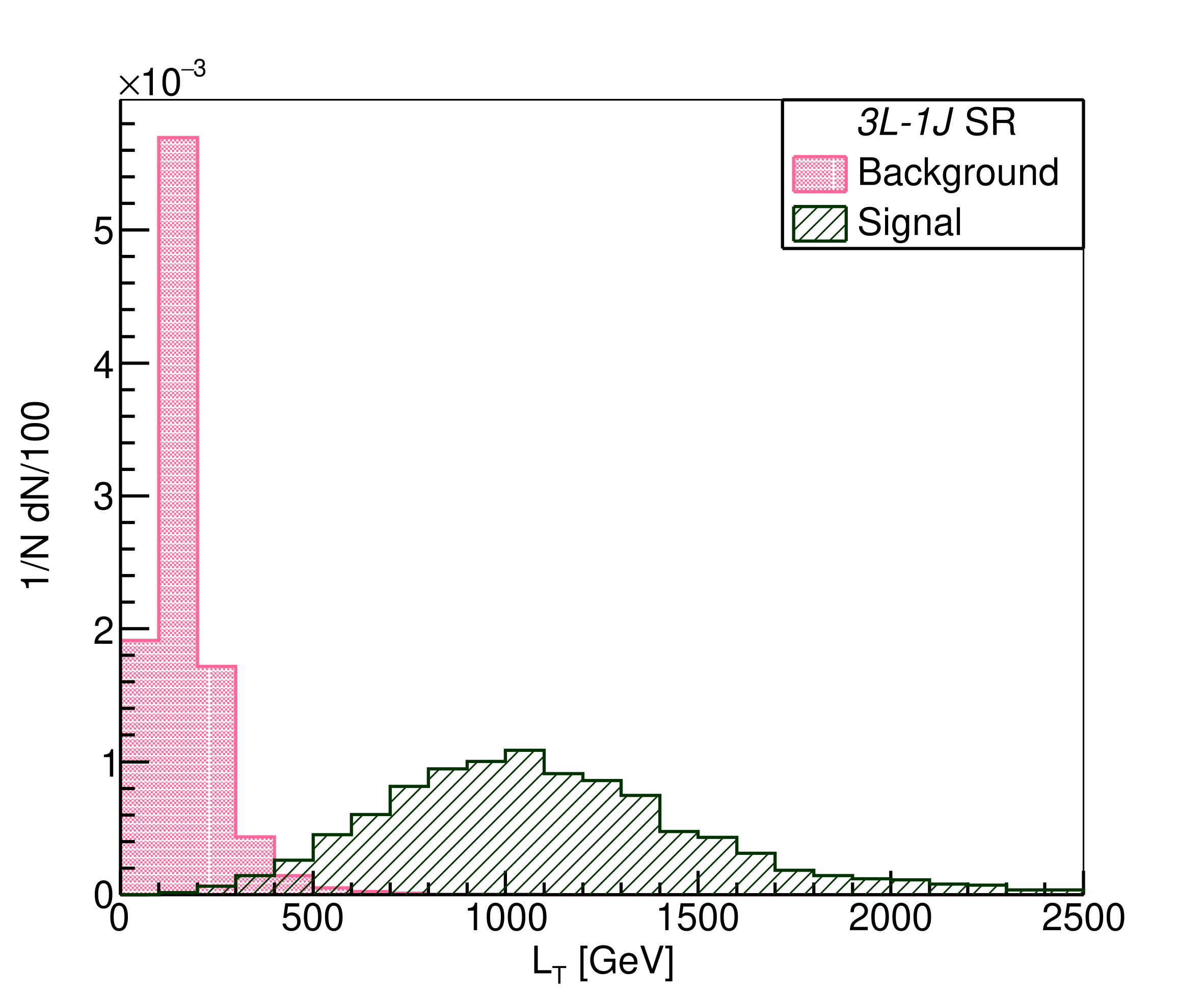}
\includegraphics[width=0.32\columnwidth]{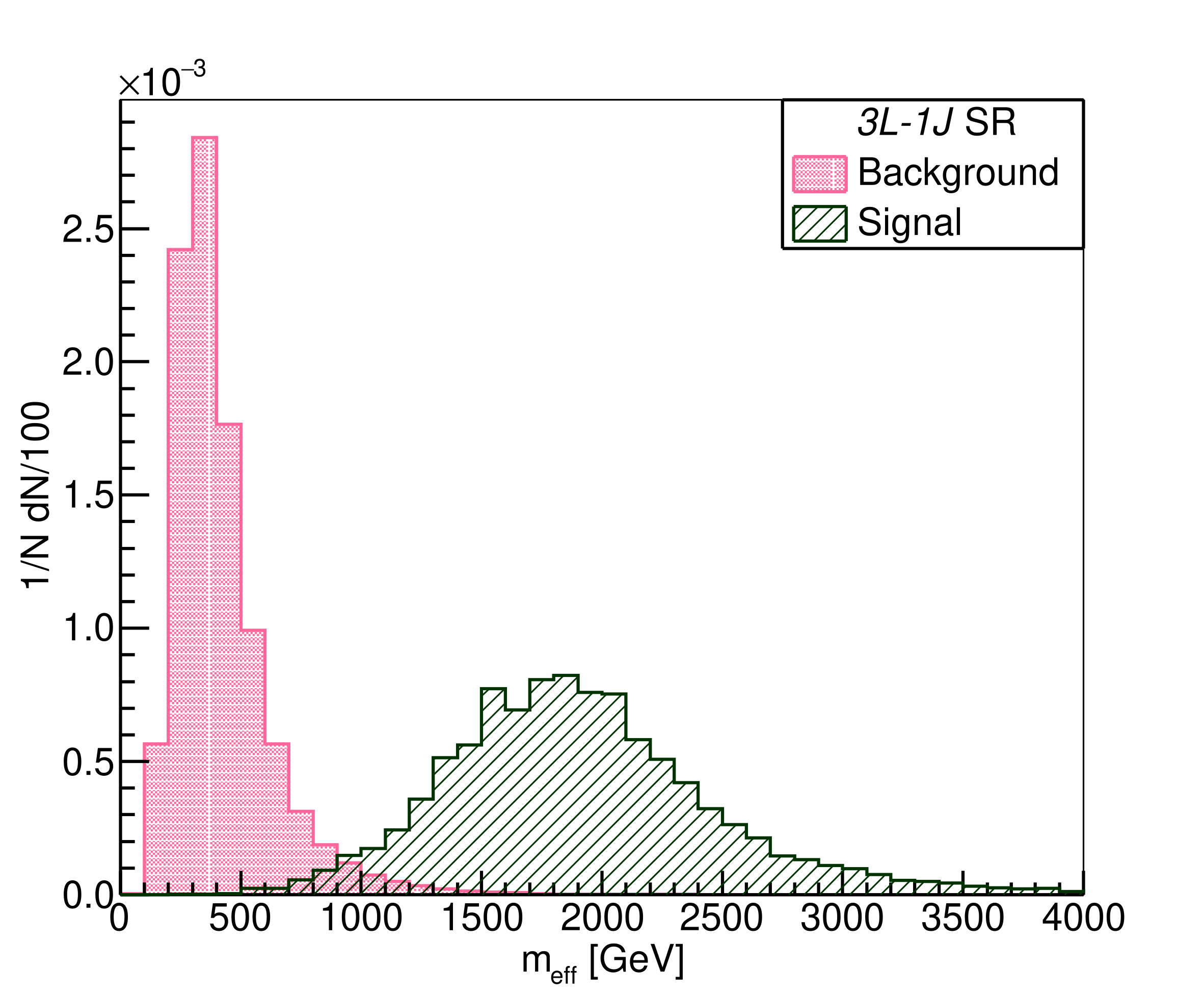}
\caption{Normalised kinematic distributions for {\it 3L-1J} events for {\it BP1}: $M_\chi=1$ TeV and {\it BR}$(\chi^\pm \to \nu W^\pm) \approx 50\%$ in {\it Model-$3^F_1$}. Top panel: $p_T^{\rm miss}$ (left), $p_T$ of the opposite-charge lepton (middle) and $p_T$ of the leading same-charge lepton (right); bottom panel: $L_T$ (left) and $M_{\rm eff}$ (right).}
\label{fig:3L1J}
\end{figure}

In Fig.~\ref{fig:3L1J}, we display different normalised kinematic distributions for a specific benchmark point \textbf{\textit{BP1}}: $M_\chi=1$ TeV and {\it BR}$(\chi^\pm \to \nu W^\pm) \approx 50\%$ in \textbf{\textit{Model-$3^F_1$}}. The left plot in the top panel shows the missing transverse momentum $p_T^{\rm miss}$ distribution, whereas
the middle and right plots show, respectively, the $p_T$ distributions of the opposite-charge and leading same-charge leptons, $p_T(\ell^{\rm oc})$ and $p_T(\ell^{\rm sc}_0)$. In the bottom panel, we plot the $L_T$ and $M_{\rm eff}=L_T+H_T+p_T^{\rm miss}$ distributions, with $L_T(H_T)$ being the scalar sum of transverse momenta of all the
leptons (jets) satisfying the reconstruction and selection requirements in an event. We see, in Fig.~\ref{fig:3L1J}, that the kinematic distributions for the signal are much harder than those for the background. Therefore, reasonably strong cuts on $p_T^{\rm miss}$, $p_T(\ell^{\rm oc})$, $p_T(\ell^{\rm sc}_0)$, $L_T$ and $M_{\rm eff}$ would appreciably curtail the background while keeping the signal relatively unharmed. Led the way by the kinematic distributions in Fig.~\ref{fig:3L1J}, we impose the following selection cuts (in GeV):
$$\textbf{Selection~I-1~(\textit{SI-1}):}~~p_T^{\rm miss} > 100, \,\, p_T(\ell^{oc}) > 100, \,\, p_T(\ell_0^{sc}) > 100, \,\, L_T > 500 \,\, {\rm and} \,\, m_{\rm eff} > 800.$$

\begin{figure}[htb!]
\centering
\includegraphics[width=0.32\columnwidth]{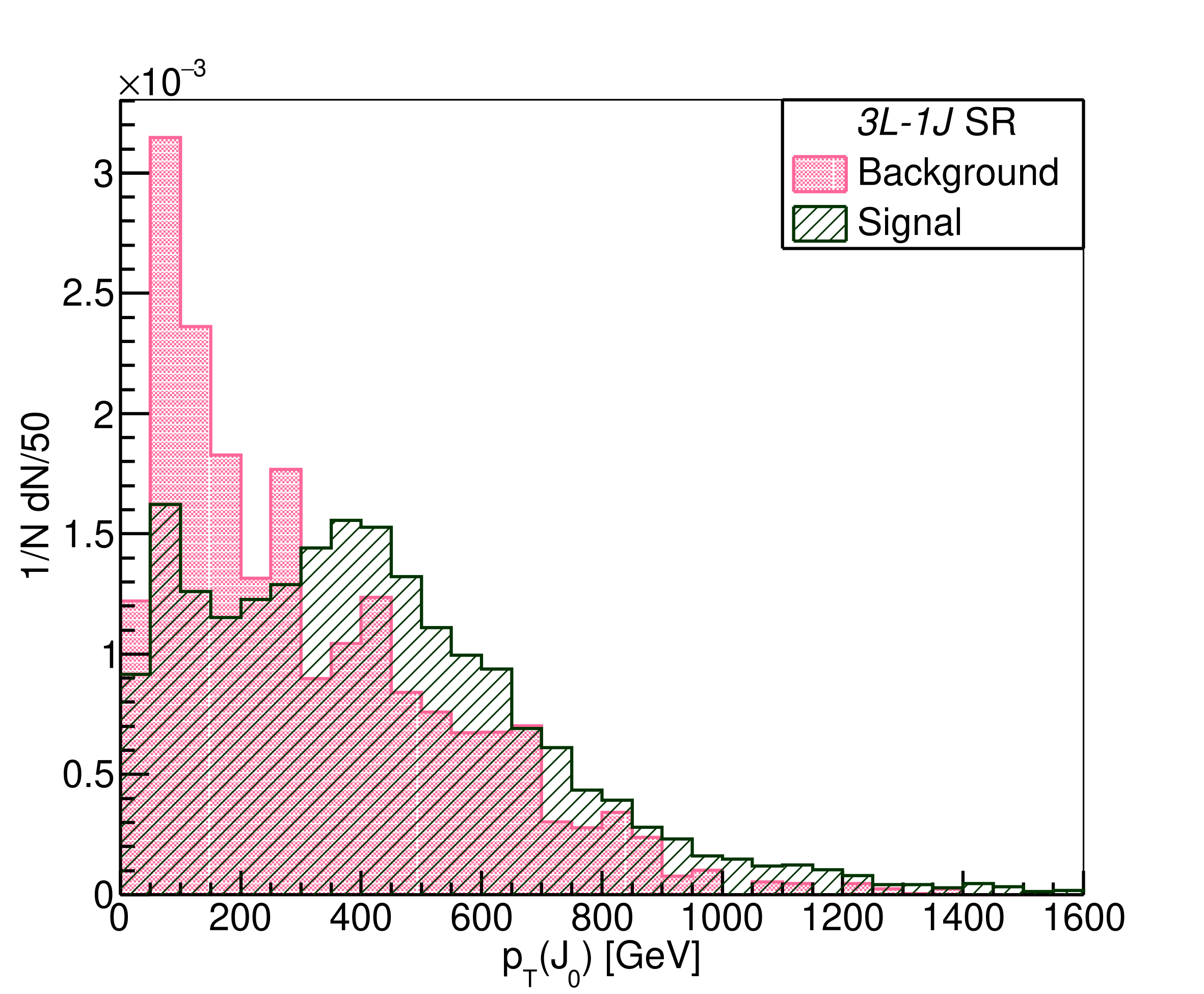}
\includegraphics[width=0.32\columnwidth]{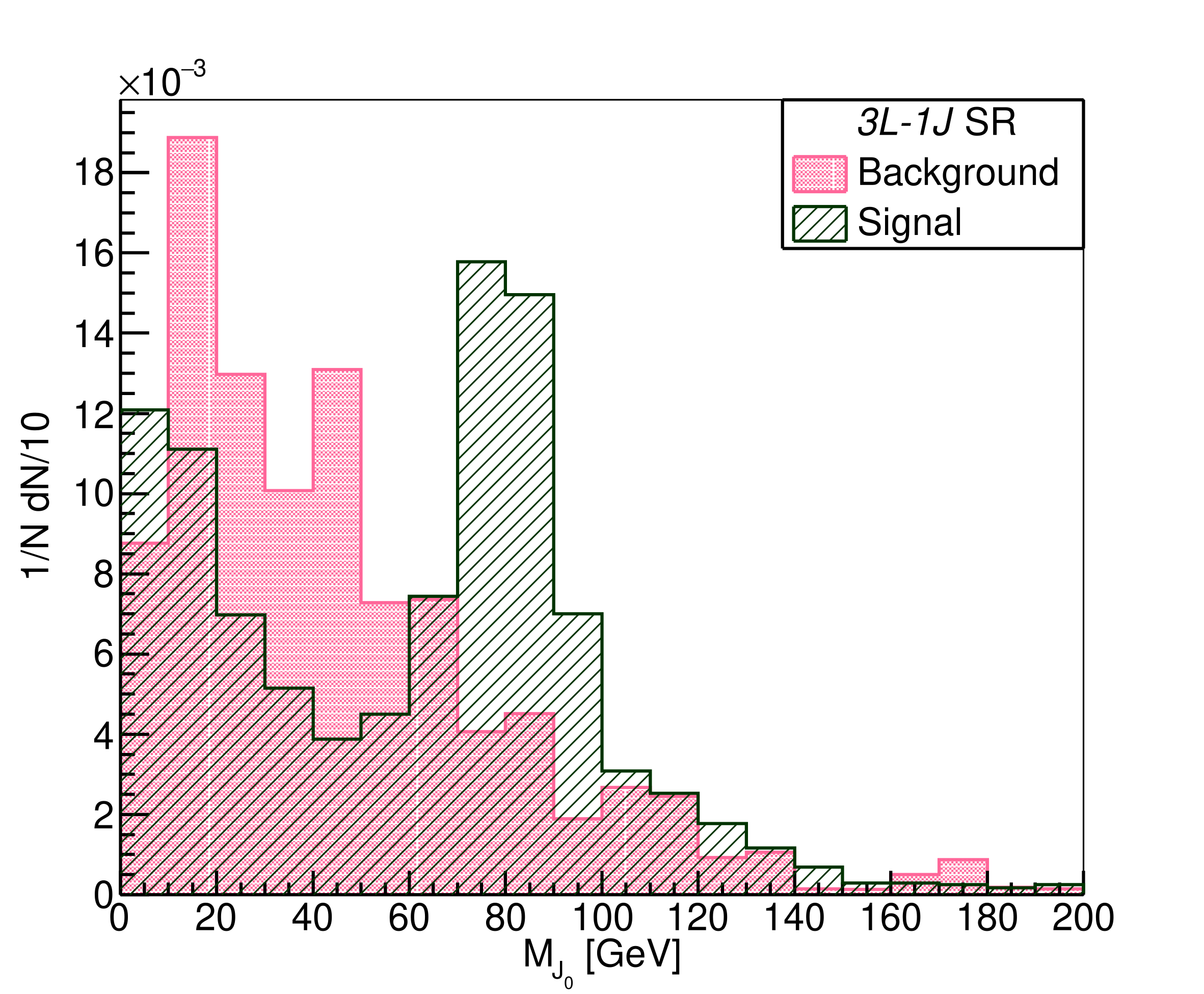}
\includegraphics[width=0.32\columnwidth]{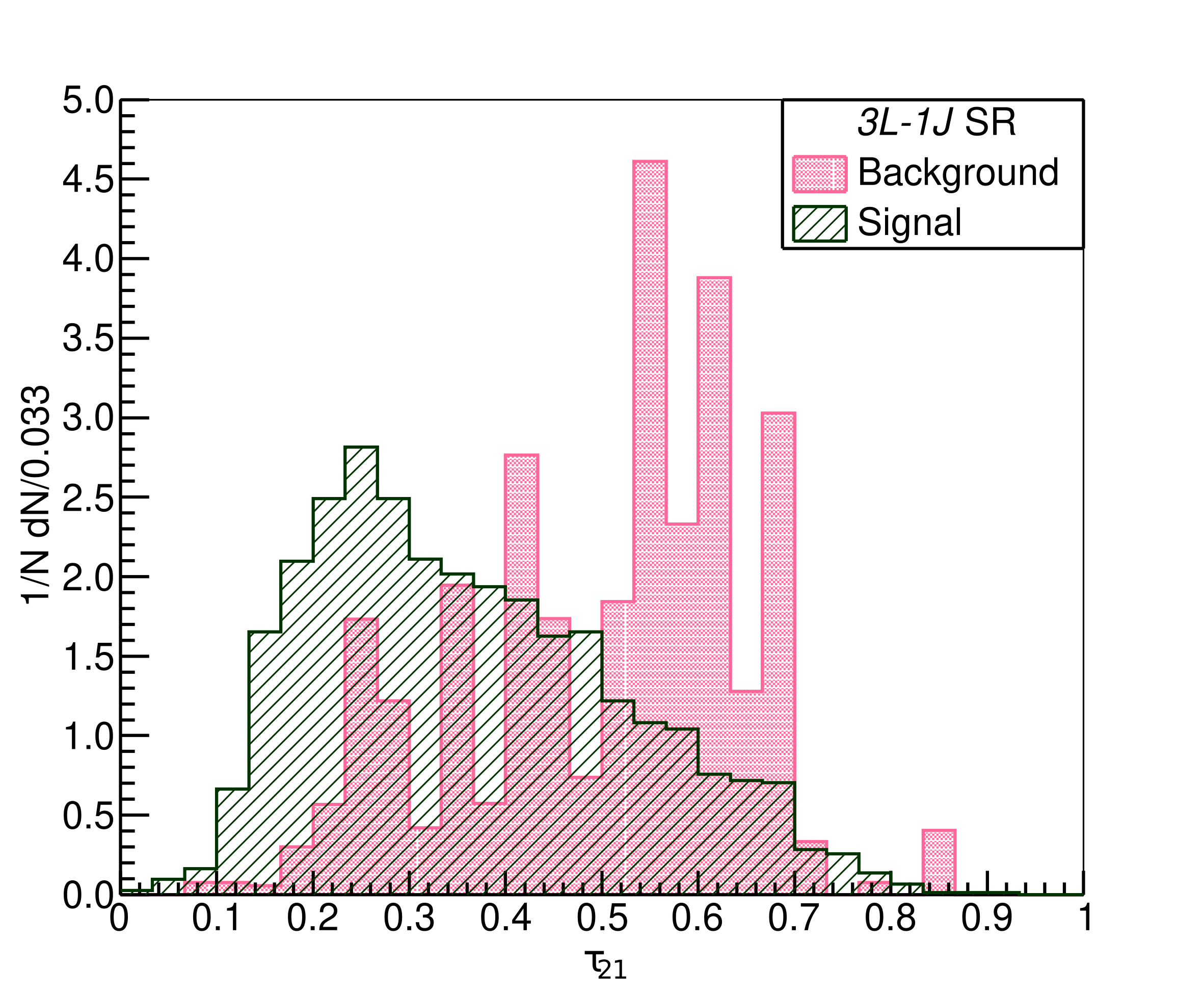}
\caption{Normalised kinematic distributions of $J_0$ for {\it 3L-1J} events for {\it BP1} in {\it Model-$3^F_1$}: $p_T(J_0)$ (left) and $M_{J_0}$ (middle) distributions for events satisfying {\it SI-1}; $\tau_{21}$ (right) distribution for events satisfying {\it SI-2}.}
\label{fig:3L1Jx}
\end{figure}

In Fig.~\ref{fig:3L1Jx} are shown a few normalised kinematic distributions pertaining to the leading fat-jet $J_0$ in such events. The left and middle panels show, respectively, the distributions of $p_T$ and mass of $J_0$, namely $p_T(J_0)$ and $M_{J_0}$, for events satisfying {\it SI-1} cuts. The $p_T(J_0)$ distribution for the background events peak at $\sim 100\,$GeV, whereas the signal events peak at a substantially larger value. However, a strong cut on $p_T(J_0)$ clearly impinges on the signal strength as well. As for the $M_{J_0}$ distribution for the background events, it is almost a monotonically falling one, barring the sharp fall at very low $M_{J_0}$ that is occasioned by the requirements on the jet energies. On the other hand, the signal, boasts a peak around 100\,GeV emanating from gauge boson decay. Taking
advantage of this, we select only those events in which at least one of the fat-jets not only has a sufficient $p_T$ but would also be identified as $W/Z/h$-fat-jet ($J_{W/Z/h}$), namely
\[
\textbf{Selection~I-2~(\textit{SI-2}):}~~p_T(J) > 200 {\rm ~GeV} \,\, {\rm and} \,\, M_{J} \in [70,140] {\rm ~GeV}.
\] 
Also displayed in the right panel is the distribution of the ratio of two subjettiness parameters, {\it viz.~}$\tau_{21} \equiv \tau_2/\tau_1$ \cite{Thaler:2010tr} of $J_0$ for events satisfying {\it SI-2}. A robust discriminant, $\tau_{21}$ can distinguish between genuine two-prong events (such as those from a boosted $W/Z/h$), which are expected to have a much lower value than pure QCD events (one-prong) or those arising from (non-boosted) top-decays. Taking inspiration from the figure, we impose a modest (and conventional) $\tau_{21}$ cut:
$$\textbf{Selection~I-3~(\textit{SI-3}):}~~\tau_{21}(J) < 0.5.$$

\subsubsection{The {\it 3L-2j} signal}
These are events very similar to the previous ({\it 3L-1J}) class, but for the fact that the jets from the bosons are not collimated enough for them to merge to a single fat-jet. Consequently, as far as the variables of Fig.~\ref{fig:3L1J} are concerned, there is relatively little difference between the corresponding
distributions. Consequently, we use the same selection cut as in the {\it 3L-1J} case:
$$\textbf{Selection~V-1~(\textit{SV-1}):}~~p_T^{\rm miss} > 100, \,\, p_T(\ell^{oc}) > 100, \,\, p_T(\ell_0^{sc}) > 100, \,\, L_T > 500 \,\, {\rm and} \,\, m_{\rm eff} > 800.$$

\begin{figure}[htb!]
\centering
\includegraphics[width=0.49\columnwidth]{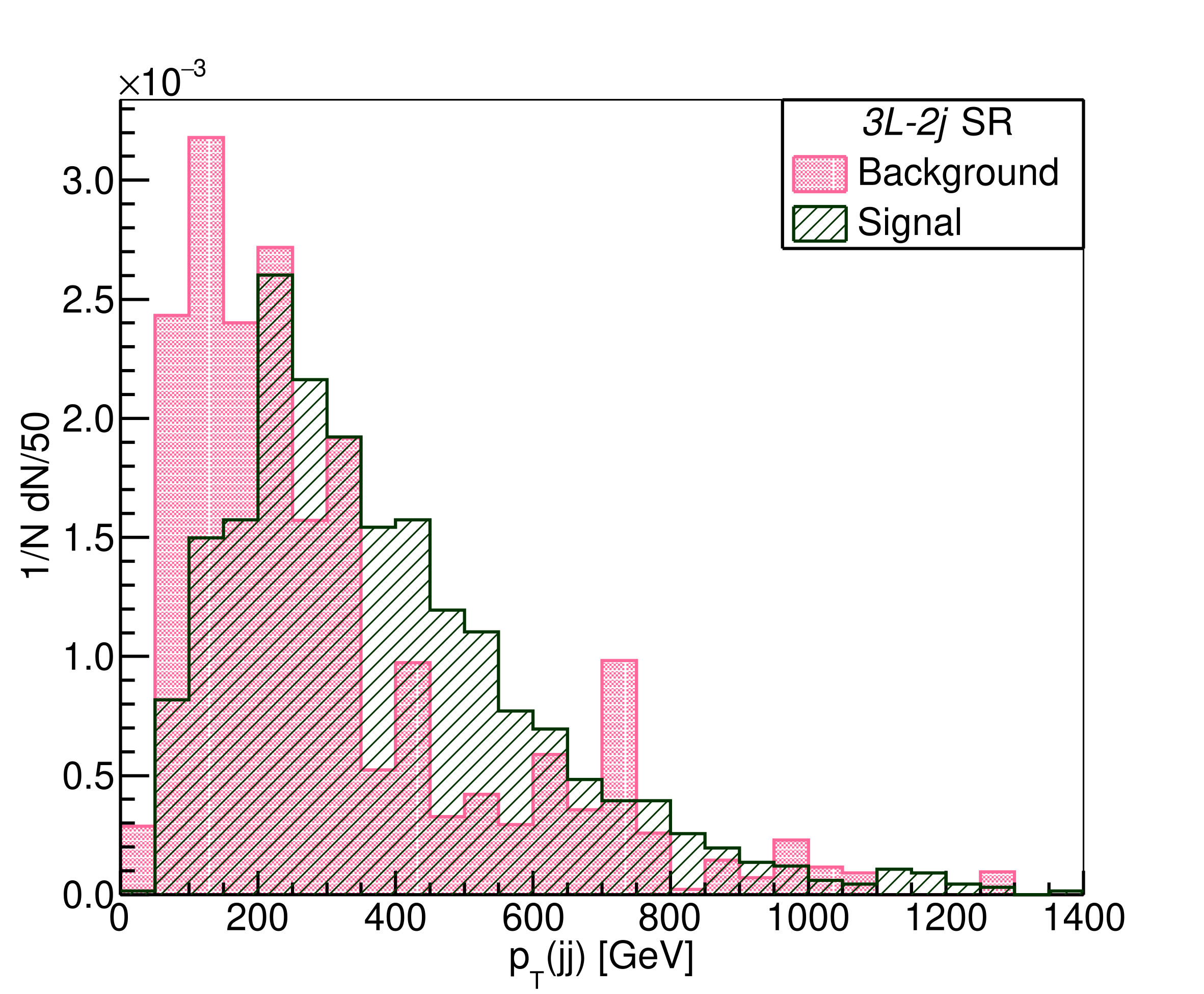}
\includegraphics[width=0.49\columnwidth]{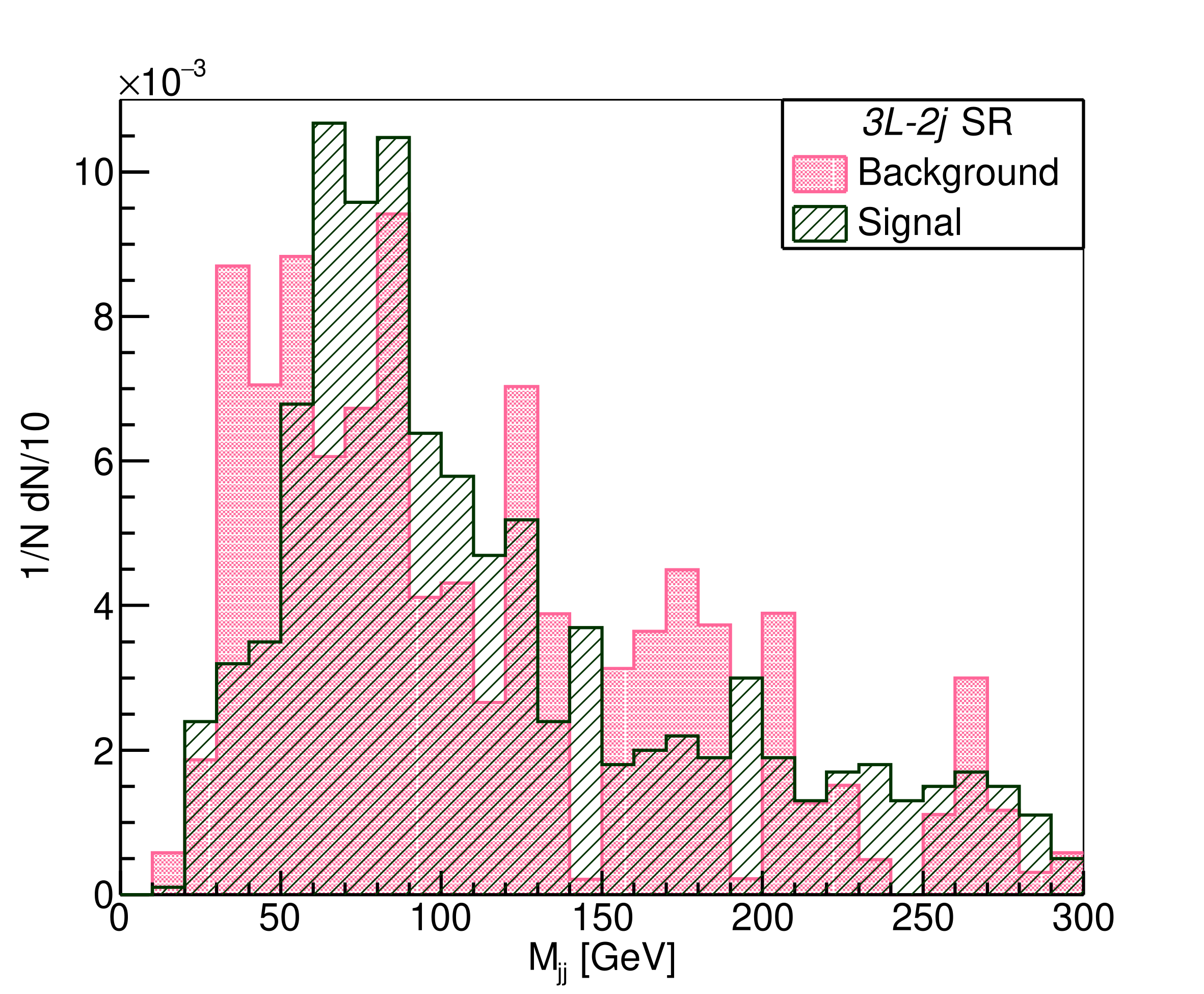}
\caption{Normalised kinematic distributions for {\it 3L-2j} events for {\it BP1} in {\it Model-$3^F_1$}: $p_T(jj)$ (left) and $M_{jj}$ (right) distributions for events satisfying {\it SI-1}.}
\label{fig:3L2j}
\end{figure}

In Fig.~\ref{fig:3L2j}, we plot the distributions for the transverse momentum (left) and invariant mass (right) of the closest two ordinary-jets system,\footnote{As mentioned earlier, the leptons and the SM bosons coming from the decays of TeV scale exotic leptons would be inordinately boosted. Thus, the jets emanating from the decaying SM bosons would be in close proximity with each other.} $p_T(jj)$ and $M_{jj}$, for the {\it 3L-2j} events satisfying {\it SI-1} cuts for {\it BP1} in {\it Model-$3^F_1$}. Because of similar structures for the background and signal distributions, it would not be possible to subjugate the background by applying selection cuts on
$p_T(jj)$ and $M_{jj}$ without harming much the signal strength. However, to enhance the signal-to-background ratio by selecting only those events in which at least one pair of the ordinary-jets emanating from gauge boson decay has a sufficient $p_T$ such that the signal would demonstrate a characteristic peak in the invariant mass distribution of a particular lepton--distinguishable from the others--and the ordinary-jets pair (see the second left plot in the top panel of Fig.~\ref{fig:SRs}), we apply the following selection cuts:\footnote{A conundrum needs to be addressed here. The selection cut {\it SV-2} improves the signal-to-noise ratio marginally but at the cost of signal strength. Therefore, whether to apply this cut or not is seemingly a dilemma. Note that, without this cut, the signal would not have a characteristic peak in the invariant mass distribution of a particular lepton and the ordinary-jets pair, thereby making the reconstruction of the exotics very challenging.}
$$\textbf{Selection~V-2~(\textit{SV-2}):}~~p_T(jj) > 200 {\rm ~GeV} \,\, {\rm and} \,\, M_{jj} \in [70,140] {\rm ~GeV}.$$

\begin{table}[htb!]
\centering
\scalebox{1.0}{
\begin{tabular}{|l|l|*{3}{p{10mm}|}|l|*{2}{p{10mm}|}}
\hline
\multirow{2}{*}{Event sample} & \multicolumn{4}{|c|}{\textit{3L-1J} SR} & \multicolumn{3}{|c|}{\textit{3L-2j} SR } \\
\cline{2-8}
& {\it S0} & {\it SI-1} & {\it SI-2} & {\it SI-3} & {\it S0} & {\it SV-1} & {\it SV-2} \\
\hline
$t\bar{t}$ & $\sim 1.1 \times 10^5$ & 177 & 41 & 11 & $\sim 6.8 \times 10^4$ & 110 & 13 \\
\hline
$WZ$ & $\sim 2.1 \times 10^4$ & 117 & 14 & 7 & 8451 & 49 & 6 \\
\hline
$t\bar{t}W$ & 2676 & 36 & 7 & 6 & 2264 & 20 & 4 \\
\hline
$t\bar{t}Z$ & 1897 & 18 & 5 & 3 & 1795 & 11 & 2 \\
\hline
Other backgrounds & $\sim 5.3 \times 10^4$ & 88 & 15 & 8 & $\sim 2.1 \times 10^4$ & 37 & 6 \\
\hline
\hline
Total background & $\sim 1.9 \times 10^5$ & 436 & 82 & 35 & $\sim 10^5$ & 227 & 31 \\
\hline
\hline
Signal & 146 & 108 & 50 & 40 & 64 & 29 & 8 \\
\hline
\end{tabular} 
}
\caption{\label{table:3L} The numbers of expected background and signal events in \textit{3L-1J} and \textit{3L-2j} SRs after passing various selection cuts for {\it BP1} in {\it Model-$3^F_1$} for 1000 fb$^{-1}$ of luminosity data at the 13 TeV LHC.}
\end{table}

Table~\ref{table:3L} displays the progression of the number of expected background and signal events (for {\it BP1} in {\it Model-$3^F_1$}) in the \textit{3L-1J} and \textit{3L-2j} SRs as subsequent selection cuts are imposed for 1000 fb$^{-1}$ of luminosity data at the 13 TeV LHC. For  the \textit{3L-1J} SR, the {\it SI-1} cut, as expected from Fig.~\ref{fig:3L1J}, turns out to be efficacious in vanquishing the background without impinging much on the signal strength while the subsequent cuts {\it SI-2} and {\it SI-3}, in particular {\it SI-2}, enhance(s) the signal-to-background ratio appreciably. Similarly, for the \textit{3L-2j} SR, the {\it SV-1} cut subdues the background while keeping almost half of the signal events, and the {\it SV-2} cut further improves the signal-to-background ratio.

\subsubsection{The other signal profiles}
As for the remaining signals discussed in Sec.~\ref{sec:object}, the distributions (and, hence, the discussions) are very similar. We do not discuss these in detail for brevity's sake but to summarise all the SR-specific selection cuts in Table~\ref{table:cuts}. The cut flows of expected background and signal events in \textit{SSD-1J} and \textit{SSD-2j} SRs are shown in Table~\ref{table:SSD}, while those for \textit{OSD-1J} and \textit{OSD-2J} SRs are shown in Table~\ref{table:OSD}.

\begin{table}[htb!]
\centering
\scalebox{0.75}{
\begin{tabular}{|p{2.65cm}|p{17cm}|}
\hline
\hline
\multicolumn{2}{|c|}{\textbf{\textit{3L-1J} SR}} \\
\hline
\textit{SI-1} & $p_T^{\rm miss} > 100 {\rm ~GeV}, \,\, p_T(\ell^{\rm oc}) > 100 {\rm ~GeV}, \,\, p_T(\ell_0^{\rm sc}) > 100 {\rm ~GeV}, \,\, L_T > 500 {\rm ~GeV} \,\, {\rm and} \,\, m_{\rm eff} > 800 {\rm ~GeV}$ \\
\hline
\textit{SI-2} and \textit{SI-3} & at least one fat-jet with $p_T(J) > 200 {\rm ~GeV}, M_{J} \in [70,140] {\rm ~GeV}$ and $\tau_{21}(J) < 0.5$ \\
\hline
\hline
\multicolumn{2}{|c|}{\textbf{\textit{SSD-1J} SR}} \\
\hline
\textit{SII-1} & $p_T^{\rm miss} > 100 {\rm ~GeV}, \,\, p_T(\ell_0) > 250 {\rm ~GeV}, \,\, L_T > 400 {\rm ~GeV} \,\, {\rm and} \,\, m_{\rm eff} > 700 {\rm ~GeV}$ \\
\hline
\textit{SII-2} and \textit{SII-3} & at least one fat-jet with $p_T(J) > 200 {\rm ~GeV}, M_{J} \in [70,140] {\rm ~GeV}$ and $\tau_{21}(J) < 0.5$ \\
\hline
\hline
\multicolumn{2}{|c|}{\textbf{\textit{OSD-1J} SR ($p_T^{\rm miss} > 100 {\rm ~GeV}$)}} \\
\hline
\textit{SIII-1} & $M_{\ell \ell} > 100 {\rm ~GeV}, \,\, p_T(\ell_0) > 300 {\rm ~GeV}, \,\, p_T(\ell_1) > 100 {\rm ~GeV}, \,\, L_T > 500 {\rm ~GeV}, \,\, H_T > 250 {\rm ~GeV} \,\, {\rm and} \,\, m_{\rm eff} > 800 {\rm ~GeV}$ \\
\hline
\textit{SIII-2} and \textit{SIII-3} & exactly one fat-jet with $p_T(J) > 200 {\rm ~GeV}, M_{J} \in [70,140] {\rm ~GeV}$ and $\tau_{21}(J) < 0.5$ \\
\hline
\hline
\multicolumn{2}{|c|}{\textbf{\textit{OSD-2J} SR}} \\
\hline
\textit{SIV-1} & $M_{\ell \ell} > 100 {\rm ~GeV}, \,\, p_T(\ell_0) > 300 {\rm ~GeV}, \,\, p_T(\ell_1) > 100 {\rm ~GeV}, \,\, L_T > 500 {\rm ~GeV}, \,\, H_T > 250 {\rm ~GeV} \,\, {\rm and} \,\, m_{\rm eff} > 800 {\rm ~GeV}$ \\
\hline
\textit{SIV-2} and \textit{SIV-3} & exactly two fat-jets with $p_T(J) > 200 {\rm ~GeV}, M_{J} \in [70,140] {\rm ~GeV}$ and $\tau_{21}(J) < 0.5$ \\
\hline
\hline
\multicolumn{2}{|c|}{\textbf{\textit{3L-2j} SR}} \\
\hline
\textit{SV-1} & $p_T^{\rm miss} > 100 {\rm ~GeV}, \,\, p_T(\ell^{\rm oc}) > 100 {\rm ~GeV}, \,\, p_T(\ell_0^{\rm sc}) > 100 {\rm ~GeV}, \,\, L_T > 500 {\rm ~GeV} \,\, {\rm and} \,\, m_{\rm eff} > 800 {\rm ~GeV}$ \\
\hline
\textit{SV-2} & at least one ordinary dijet with $p_T(jj) > 200 {\rm ~GeV} \,\, {\rm and} \,\, M_{jj} \in [70,140] {\rm ~GeV}$ \\
\hline
\hline
\multicolumn{2}{|c|}{\textbf{\textit{SSD-2j} SR}} \\
\hline
\textit{SVI-1} & $p_T^{\rm miss} > 100 {\rm ~GeV}, \,\, p_T(\ell_0) > 250 {\rm ~GeV}, \,\, L_T > 500 {\rm ~GeV} \,\, {\rm and} \,\, m_{\rm eff} > 700 {\rm ~GeV}$ \\
\hline
\textit{SVI-2} & at least one ordinary dijet with $p_T(jj) > 200 {\rm ~GeV} \,\, {\rm and} \,\, M_{jj} \in [70,140] {\rm ~GeV}$ \\
\hline
\hline
\end{tabular} 
}
\caption{\label{table:cuts} Summary of the SR-specific selection cuts. The {\it OSD-1J} and {\it OSD-2J} SRs become mutually exclusive after all the selection cuts are imposed.}
\end{table}

\begin{table}[htb!]
\centering
\scalebox{1.0}{
\begin{tabular}{|l|l|*{3}{p{10mm}|}|l|*{2}{p{10mm}|}}
\hline
\multirow{2}{*}{Event sample} & \multicolumn{4}{|c|}{\textit{SSD-1J} SR} & \multicolumn{3}{|c|}{\textit{SSD-2j} SR } \\
\cline{2-8}
& {\it S0} & {\it SII-1} & {\it SII-2} & {\it SII-3} & {\it S0} & {\it SVI-1} & {\it SVI-2} \\
\hline
$t\bar{t}$ & $\sim 1.2 \times 10^5$ & 271 & 104 & 63 & $\sim 8.9 \times 10^4$ & 97 & 13 \\
\hline
$t\bar{t}W$ & $\sim 1.5 \times 10^4$ & 138 & 48 & 30 & $\sim 1.5 \times 10^4$ & 64 & 13 \\
\hline
$WWW$ & $\sim 1.2 \times 10^4$ & 93 & 40 & 28 & $\sim 10^4$ & 31 & 4 \\
\hline
$WZ$ & $\sim 5.6 \times 10^4$ & 160 & 52 & 27 & $\sim 2.7 \times 10^4$ & 69 & 14 \\
\hline
Other backgrounds & $\sim 8.4 \times 10^4$ & 145 & 52 & 31 & $\sim 5.2 \times 10^4$ & 62 & 9 \\
\hline
\hline
Total background & $\sim 2.9 \times 10^5$ & 807 & 296 & 179 & $\sim 1.9 \times 10^5$ & 323 & 53 \\
\hline
\hline
Signal & 141 & 90 & 51 & 41 & 79 & 38 & 9 \\
\hline
\end{tabular} 
}
\caption{\label{table:SSD} The numbers of expected background and signal events in \textit{SSD-1J} and \textit{SSD-2j} SRs after passing various selection cuts for {\it BP1} in {\it Model-$3^F_1$} for 1000 fb$^{-1}$ of luminosity data at the 13 TeV LHC.}
\end{table}

\begin{table}[htb!]
\centering
\scalebox{0.92}{
\begin{tabular}{|l|l|*{3}{p{10mm}|}|l|l|*{2}{p{10mm}|}}
\hline
\multirow{2}{*}{Event sample} & \multicolumn{4}{|c|}{\textit{OSD-1J} SR} & \multicolumn{4}{|c|}{\textit{OSD-2J} SR } \\
\cline{2-9}
& {\it S0} & {\it SIII-1} & {\it SIII-2} & {\it SIII-3} & {\it S0} & {\it SIV-1} & {\it SIV-2} & {\it SIV-3} \\
\hline
$t\bar{t}$ & $\sim 3.1 \times 10^6$ & 2653 & 719 & 265 & $\sim 10^7$ & 4317 & 42 & 3 \\
\hline
$WW$ & $\sim 1.1 \times 10^6$ & 738 & 240 & 89 & $\sim 4.2 \times 10^6$ & 1070 & 5 & 1 \\
\hline
$\gamma^*/Z$ & $\sim 1.6 \times 10^7$ & 503 & 168 & 64 & $\sim 5.1 \times 10^8$ & 6427 & 76 & 17 \\
\hline
$WZ$ & 7781 & 89 & 34 & 14 & $\sim 5.4 \times 10^4$ & 393 & 10 & 1 \\
\hline
Other backgrounds & $\sim 3.2 \times 10^4$ & 199 & 96 & 50 & $\sim 10^5$ & 400 & 15 & 6 \\
\hline
\hline
Total background & $\sim 2.0 \times 10^7$ & 4182 & 1257 & 482 & $\sim 5.3 \times 10^8$ & $\sim 1.3 \times 10^4$ & 148 & 28 \\
\hline
\hline
Signal & 339 & 194 & 149 & 92 & 382 & 269 & 63 & 39 \\
\hline
\end{tabular} 
}
\caption{\label{table:OSD} The numbers of expected background and signal events in \textit{OSD-1J} and \textit{OSD-2J} SRs after passing various selection cuts for {\it BP1} in {\it Model-$3^F_1$} for 1000 fb$^{-1}$ of luminosity data at the 13 TeV LHC.}
\end{table}
 
\begin{figure}[htb!]
\centering
\includegraphics[width=0.5\columnwidth]{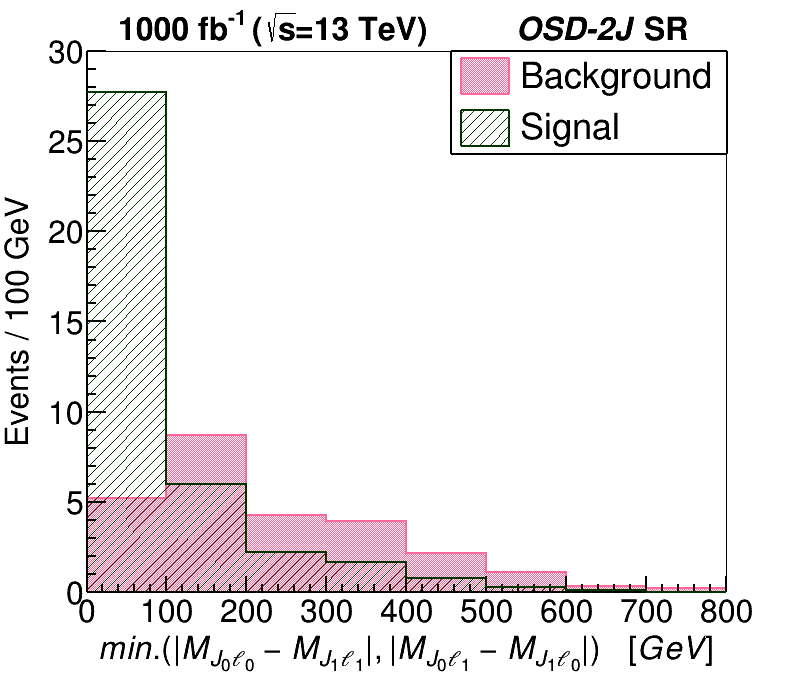}
\caption{Distributions of the difference between the two lepton-fat-jet invariant masses for the {\it OSD-2J} SR for \textit{BP1}: $M_\chi=1$ TeV and {\it BR}$(\chi^\pm \to \nu W^\pm) \approx 50\%$ in {\it Model-$3^F_1$}. The events are weighted for 1000 fb$^{-1}$ at the 13 TeV LHC.}
\label{fig:inv_mass_diff}
\end{figure}

To enhance the sensitivity of this search, the signal regions are further classified into several independent search bins using a primary kinematic discriminant between the signal and background:

\begin{itemize}
\item The {\it 3L-1J} ({\it 3L-2j}) SR uses the invariant mass of the opposite-charge lepton and fat-jet (ordinary dijet) system, $M_{J\ell^{oc}}$ ($M_{jj\ell^{oc}}$), as the primary kinematic discriminant. For the signal events, this invariant mass distribution would peak at the mass of the exotic lepton.

\item For the {\it SSD-1J} and {\it SSD-2j} SRs, the exotic lepton mass can not be reconstructed from any invarint mass distribution and, thus, we use $L_T+p_T^{\rm miss}$ as primary kinematic discriminant.

\item For the {\it OSD-1J} SR, we use $M_{J \ell_0}$ as primary kinematic discriminant.

\item For the {\it OSD-2J} SR, the kinematic reconstruction of both the exotics is possible. Since the mass difference between dissimilar exotics is small (see Table \ref{table:mass-splitting}), we may safely assume the pairing with the smaller mass difference to be the correct one. For the signal events, this mass difference is expected to be dominated by the uncertainties inherent in jet reconstruction, while for the background events, large additional contributions accrue from the underlying hard scattering process itself. This difference is exemplified by the distributions depicted in Fig.~\ref{fig:inv_mass_diff}, and their very shapes induce us to impose the final selection cut, namely restrict this difference to be less than 100~GeV. As the final kinematic discriminant, we use the average of the two lepton-fat-jet invariant masses computed from the leading particles that correspond to the smaller difference. 
In other words, if $|M_{J_0 \ell_0}-M_{J_1 \ell_1}| < |M_{J_1 \ell_0}-M_{J_0 \ell_1}|$, then the discriminant is $M_{J \ell}^{\rm avg}=(M_{J_0 \ell_0}+M_{J_1 \ell_1})/2$, else it is $M_{J \ell}^{\rm avg}=(M_{J_0 \ell_1}+M_{J_1 \ell_0})/2$.
\end{itemize}

Clearly, except in the second ({\it SSD-1J} and {\it SSD-2j}) case, these invariant mass distributions could be used to reconstruct the exotic leptons kinematically. Each
SR is divided into ten independent search bins in the [0:2000] GeV range\footnote{The range [0:2000] GeV is chosen to maximise the sensitivity of the present search for the exotics of 1-2 TeV masses. However, for the exotics heavier than 2 TeV, a broader range (say [1000:3000] GeV) would be beneficent in reconstructing the exotics.} for the corresponding discriminant. The width of the bins is chosen to ensure smooth and monotonic behaviour of the expected background while retaining sensitivity to the simplified models. The overflow events are contained in the last bin for each signal region. This binning scheme yields a total of 60 statistically independent search bins (see Table~\ref{table:binning}).

\begin{table}[htb!]
\centering
\begin{tabular}{|l c c c|}
\hline
Label & Kinematic discriminant & Range (GeV) &  Number of bins 
\\
\hline
{\it 3L-1J} & $M_{J\ell^{oc}}$ & $[0,2000]$ & 10
\\
{\it 3L-2j} & $M_{jj\ell^{oc}}$ & $[0,2000]$ & 10
\\
{\it SSD-1J} & $L_T$+$p_T^{\rm miss}$ & $[0,2000]$ & 10
\\
{\it SSD-2j} & $L_T$+$p_T^{\rm miss}$ & $[0,2000]$ & 10
\\
{\it OSD-1J} & $M_{J \ell_0}$ & $[0,2000]$ & 10
\\
{\it OSD-2J} & $M_{J \ell}^{\rm avg}$ & $[0,2000]$ & 10
\\
\hline
\end{tabular}
\caption{Binning scheme of different signal regions.}
\label{table:binning}
\end{table}

\begin{figure}[htb!]
\centering
\includegraphics[width=0.32\columnwidth]{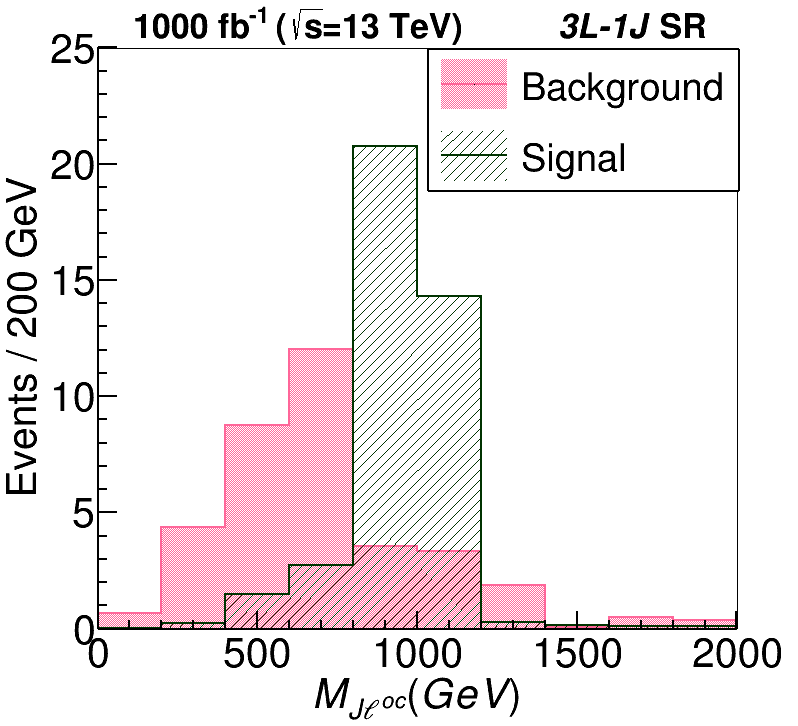}
\includegraphics[width=0.32\columnwidth]{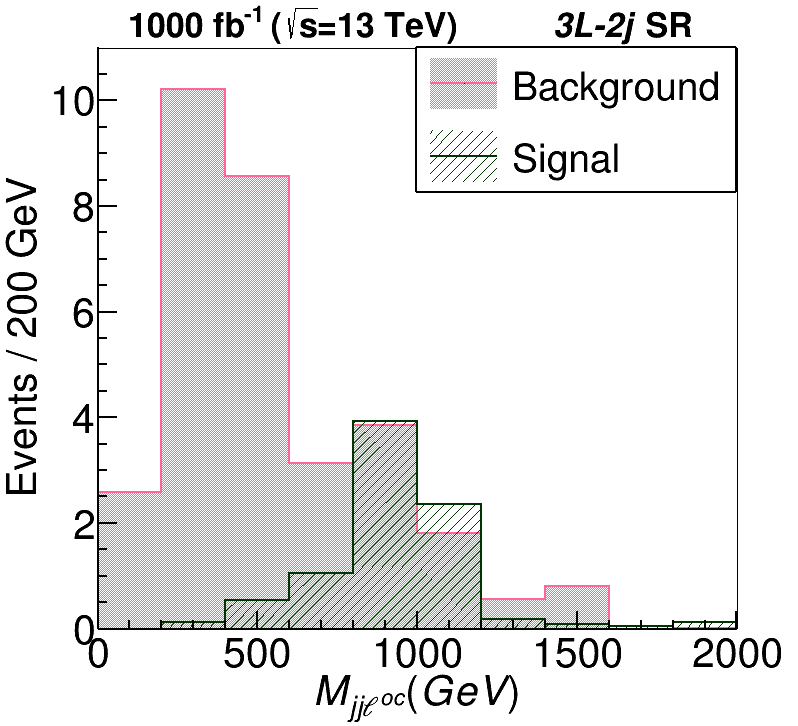}
\includegraphics[width=0.32\columnwidth]{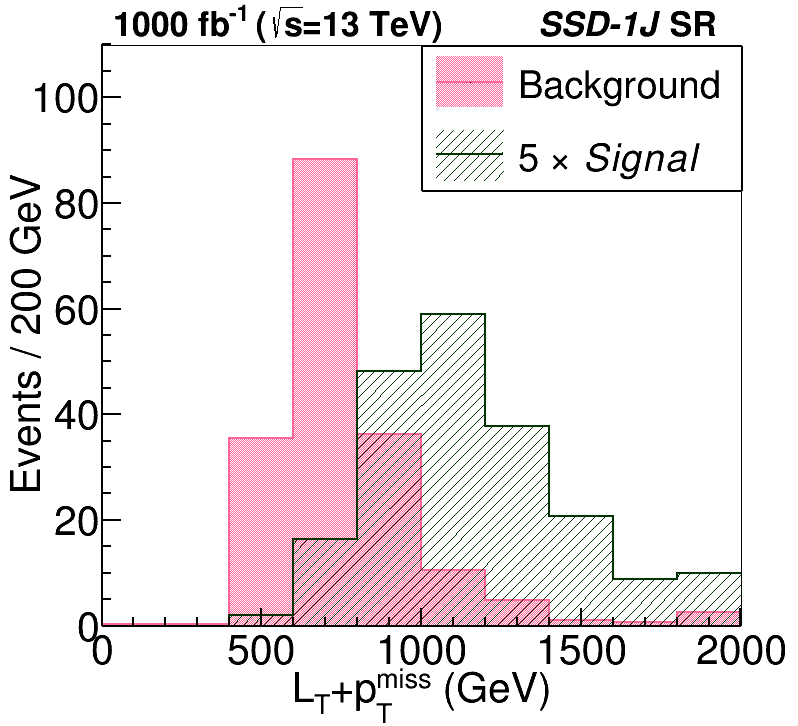}
\includegraphics[width=0.32\columnwidth]{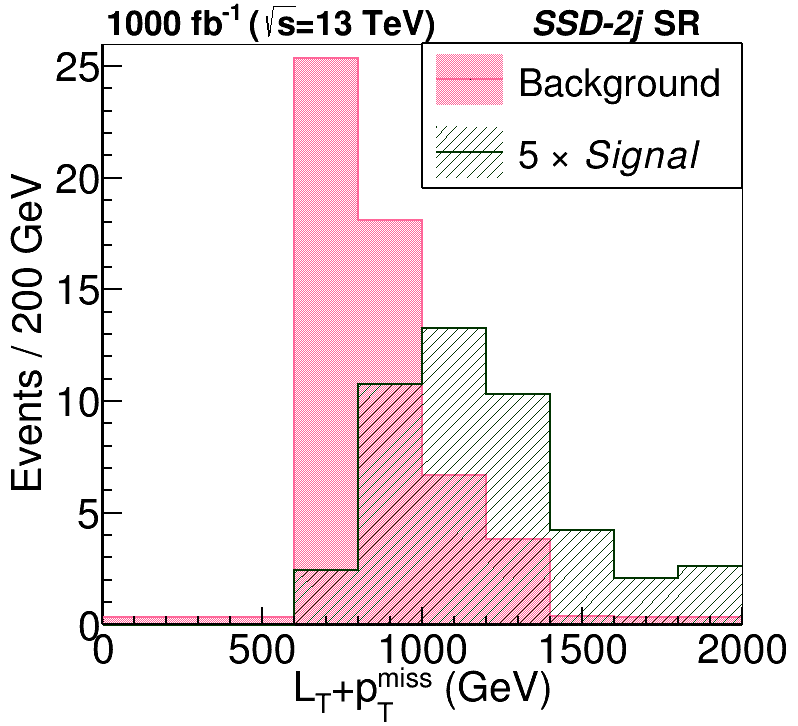}
\includegraphics[width=0.32\columnwidth]{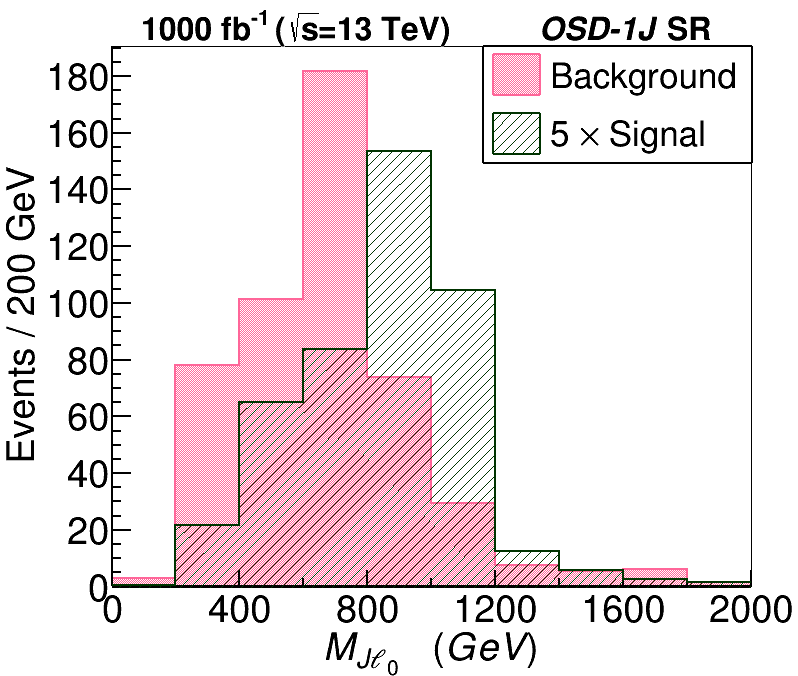}
\includegraphics[width=0.32\columnwidth]{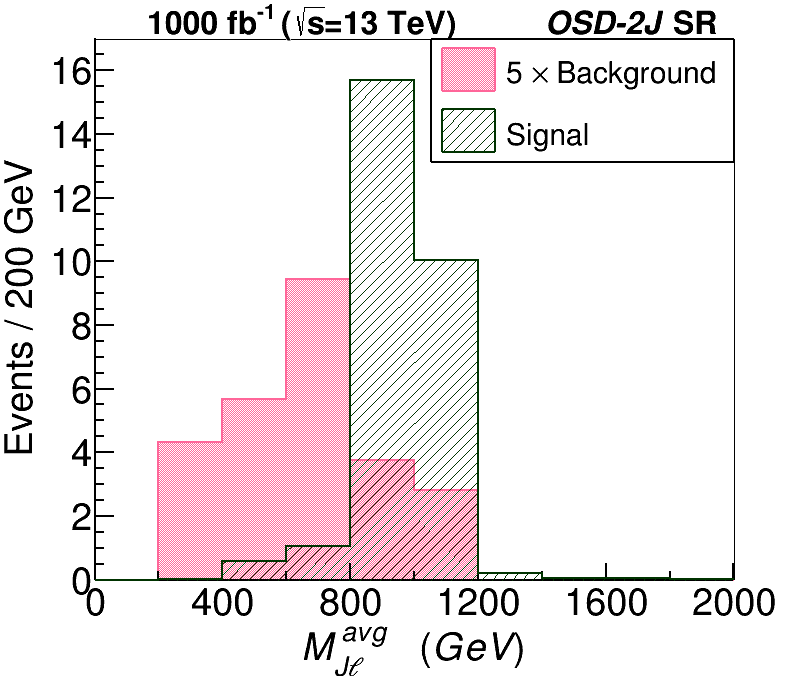}
\caption{Distributions of the primary kinematic discriminants for all the SRs for \textit{BP1}: $M_\chi=1$ TeV and {\it BR}$(\chi^\pm \to \nu W^\pm) \approx 50\%$ in {\it Model-$3^F_1$}. The events are weighted for 1000 fb$^{-1}$ at the 13 TeV LHC.}
\label{fig:SRs}
\end{figure}

\noindent
In Fig.~\ref{fig:SRs}, we show the distributions of the primary kinematic discriminants for both signal and background events passing all the selection cuts for all the SRs for \textit{BP1}: $M_\chi=1$ TeV and {\it BR}$(\chi^\pm \to \nu W^\pm) \approx 50\%$ in {\it Model-$3^F_1$}. The events are weighted for 1000 fb$^{-1}$ luminosity at the 13 TeV LHC. It is evident from Fig.~\ref{fig:SRs} that the exotics can be kinematically reconstructed from the distributions of the primary discriminants in the {\it 3L-1J}, {\it 3L-2j}, {\it OSD-1J} and {\it OSD-2J} SRs. As expected, these invariant mass distributions peak near 1000 GeV and thereby reconstruct the exotic mass in the {\it Model-$3^F_1$}.

\subsection{Discovery reach for the simplified models}
To estimate the discovery reach of the search strategies presented above, we use a hypothesis tester named \texttt{Profile Likelihood Number Counting Combination}, which uses a library of \texttt{C++} classes \texttt{RooFit} \cite{Verkerke:2003ir} in the \texttt{ROOT} \cite{Brun:1997pa} environment. Treating all the bins in different signal regions as independent channels for both signal and background events, the uncertainties are included via the {\it Profile Likelihood Ratio}. Without going into the intricacy of estimating the background uncertainties, we assume an overall 20\% total uncertainty on the estimated background.\footnote{The SM backgrounds usually suffer from substantial uncertainties---both experimental and theoretical. The experimental uncertainties arise from several sources such as the reconstruction, identification, isolation, trigger efficiency, energy scale and resolution of different physics objects, the luminosity measurements, the pile-up modelling, and the data-driven estimation of the misidentified lepton and charge-misidentified electron backgrounds. The theoretical uncertainties arise from the normalisation of some of the dominant irreducible backgrounds, the parton-shower modelling, the higher-order QCD corrections (renormalisation, factorisation and resummation scales), the PDF sets, and the strong coupling constant. Further, the limited statistics in the data and simulation samples entail significant uncertainty to the total background. For the typical LHC searches \cite{ATLAS:2018ghc,CMS:2019lwf,ATLAS:2020wop,ATLAS:2021xxb,CMS:2021zkl,ATLAS:2021eyc}, both the theoretical and experimental uncertainties are $\mathcal{O}(10)\%$ yielding a total uncertainty of less than 20\%.}Further, to ensure robustness in statistical interpretations, we replace the less than one per-bin expected background yield at 3000 fb$^{-1}$ with one background yield.

\begin{figure}[htb!]
\centering
\includegraphics[width=0.5\columnwidth]{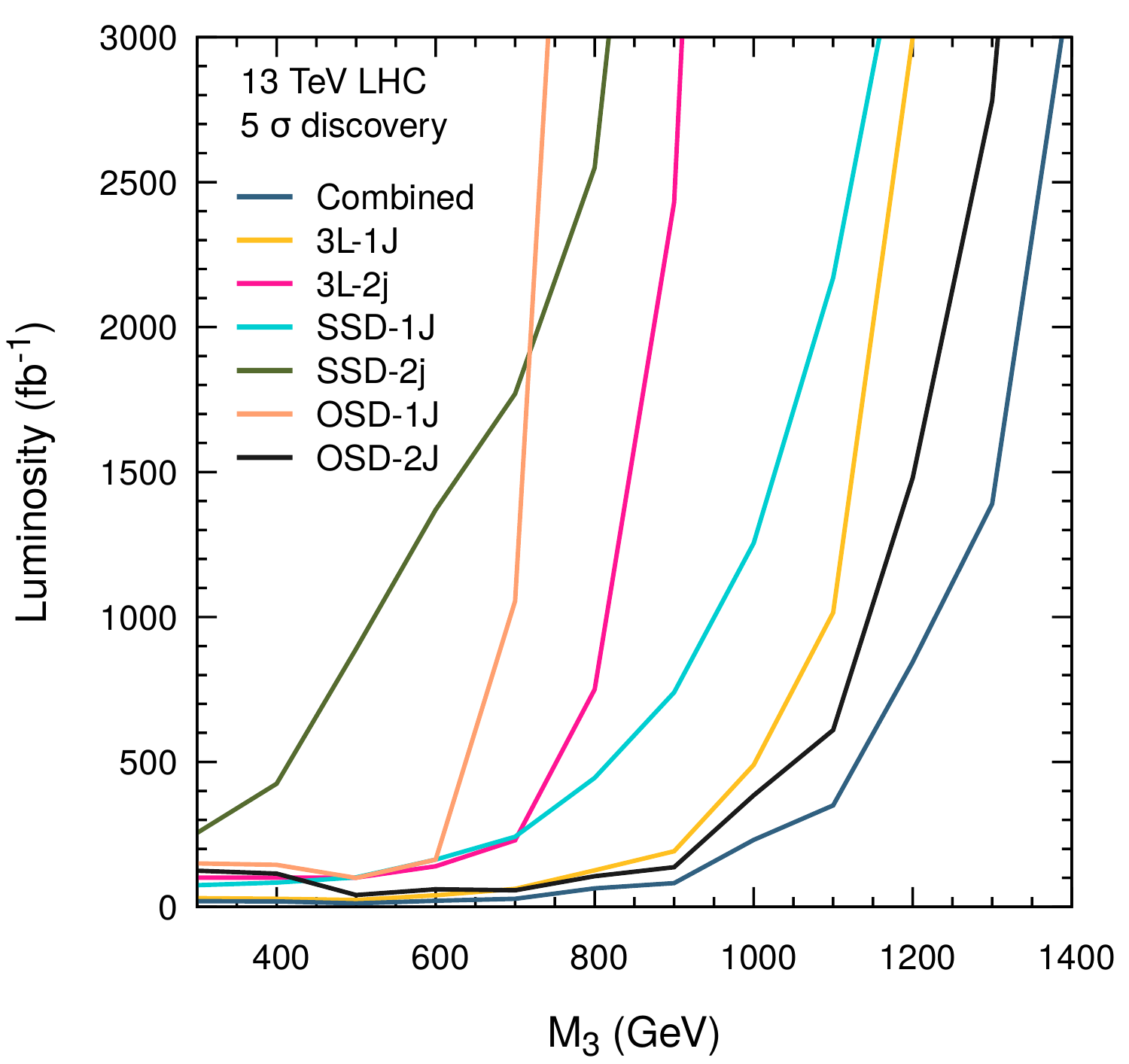}
\caption{Required Luminosity for the $5\sigma$ discovery for the {\it Model-$3^F_1$} with {\it BR}$(\chi^\pm \to \nu W^\pm) \approx 50\%$ in different SRs.}
\label{fig:lumi_5s}
\end{figure}

In Fig.~\ref{fig:lumi_5s}, projections of the required luminosities---in individual channels---for a $5\sigma$ discovery of the exotics in the {\it Model-$3^F_1$} with {\it BR}$(\chi^\pm \to \nu W^\pm) \approx 50\%$ are shown as a function of the mass. Since all the SRs considered in this search are mutually exclusive, it is reasonable to combine them together, and this too is depicted in the same figure. That the {\it OSD-1J}, {\it SSD-2j} and {\it 3L-2j} SRs are the least sensitive among all, is understandable. For the {\it OSD-1J} SR, the SM background is rather large, whereas for the {\it SSD-2j} and {\it 3L-2j} SRs, the signal strengths are small.\footnote{For the {\it OSD-1J}, {\it SSD-2j} and {\it 3L-2j} SRs, the curves in Fig.~\ref{fig:lumi_5s} have sharp turns. For one, the signals are beset with comparatively large SM backgrounds for those SRs, and second, sizeable uncertainty (which we assumed to be flat 20\%) on the latter that even increasing luminosity does not help much. With the increasing amount of data at the LHC (and thus better insight into the detector response), the experimental uncertainties are expected to be reduced significantly. However, a precise estimation of the same is beyond the realm of this work.}Consequently, the $5\sigma$ discovery reach for the $3^F_1$ exotics extends up to 910 GeV in these SRs. On the contrary, the {\it OSD-2J} SR is the most promising one with the discovery reach of 1310 GeV because of its small background and ability to reconstruct the mass of the exotic leptons without any ambiguity. Indeed, this contributes the maximum to the total sensitivity reach of approximately 1385 GeV reached by combining all
the channels.

\begin{figure}[htb!]
\centering
\includegraphics[width=0.47\columnwidth]{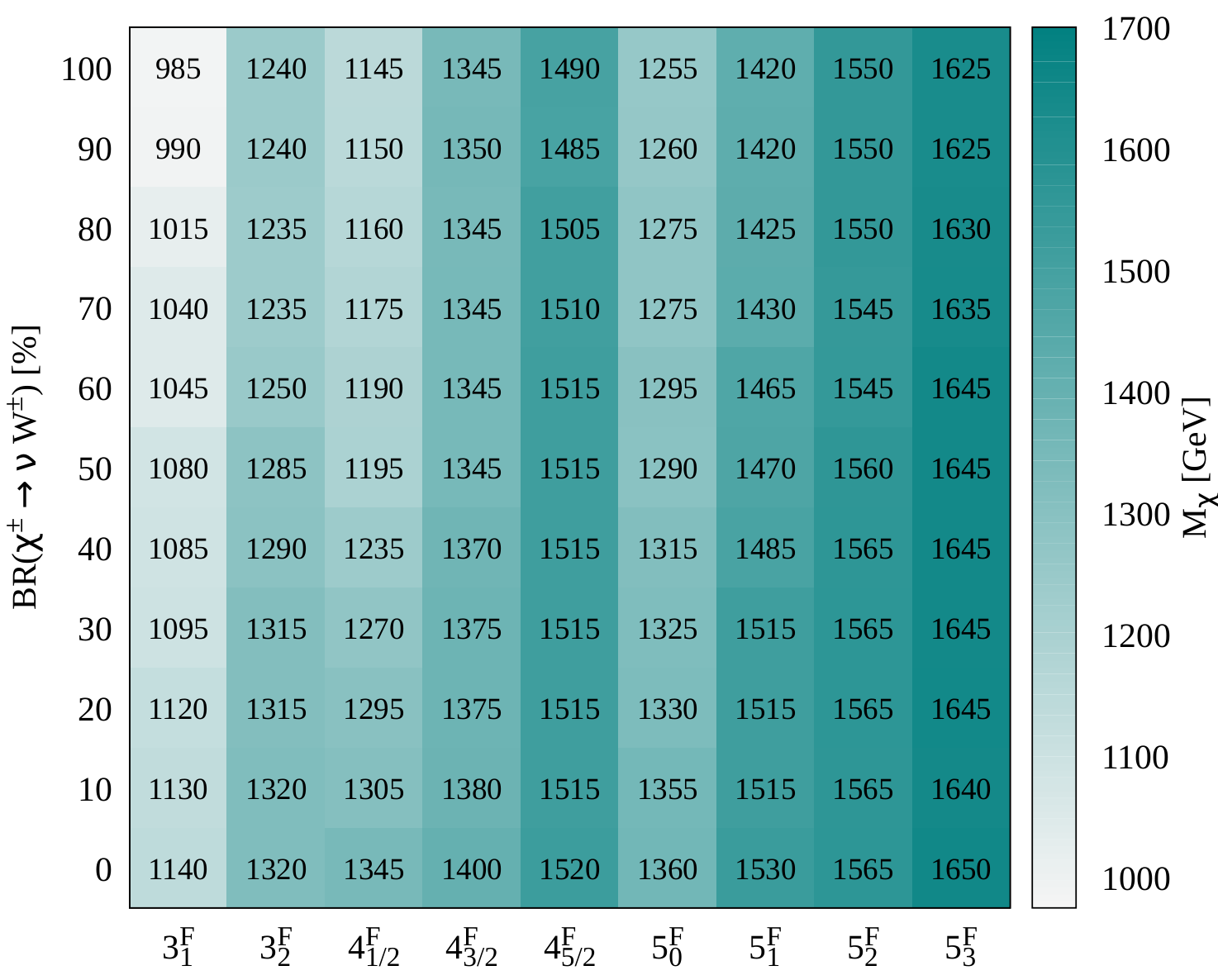} \quad
\includegraphics[width=0.47\columnwidth]{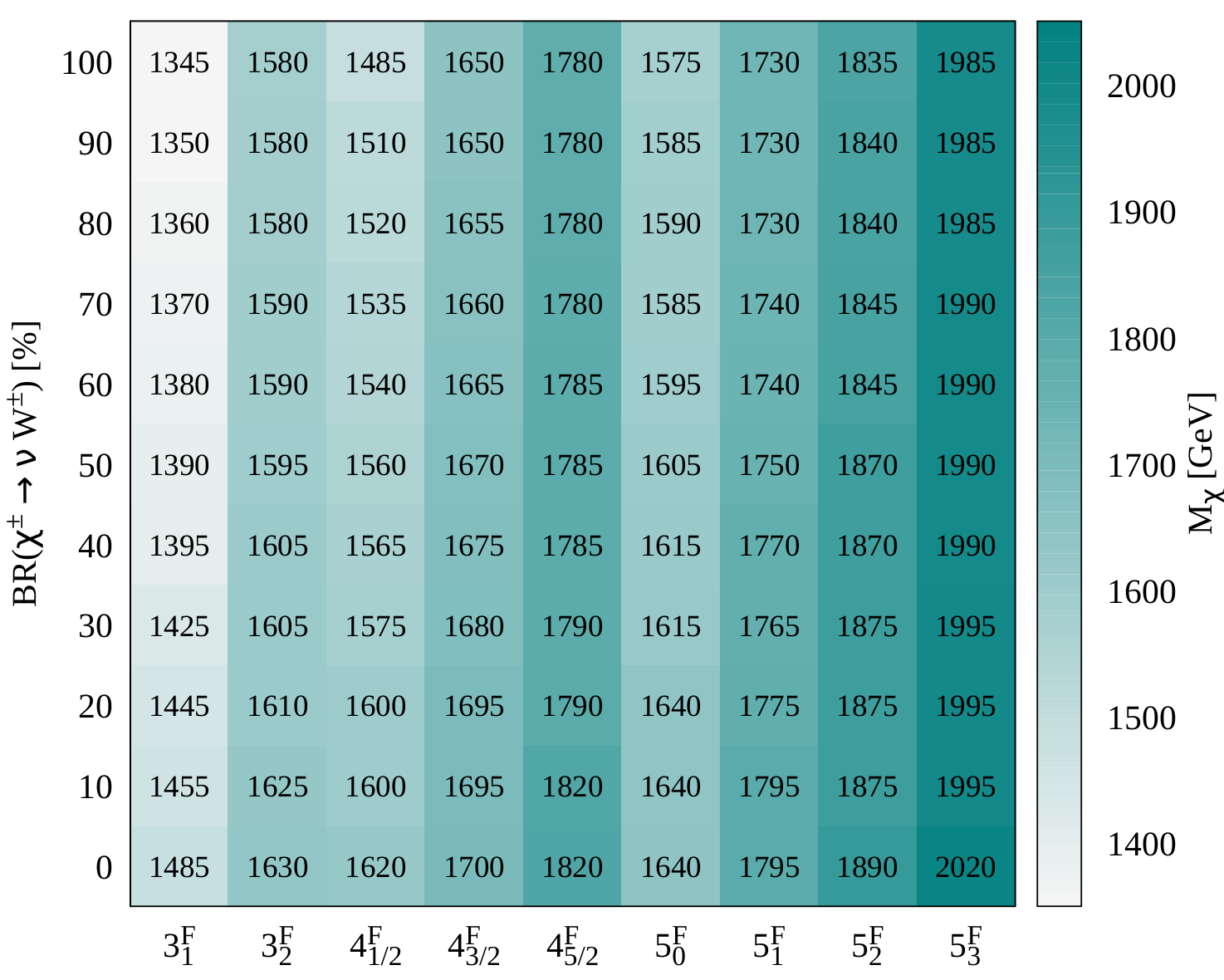}
\caption{$5\sigma$ discovery reach of exotic leptons belonging to different simplified models for different values of {\it BR}($\chi^\pm \to \nu W^\pm$) for 300 (left) and 3000 (right) fb$^{-1}$ luminosity of data at 13 TeV LHC.}
\label{fig:5s}
\end{figure}

Proceeding to a comparison of the different scenarios, we present, in Fig.~\ref{fig:5s}, the $5\sigma$ discovery reach, as defined by combining all the channels. The left and right plots correspond, respectively, to 300 and 3000 fb$^{-1}$ luminosity of data at 13 TeV LHC. A key input here is the {\it BR}($\chi^\pm \to \nu W^\pm$). While this branching fraction is trivially calculable if the multiplet represented the only BSM physics operative at these scales, it would depend on other parameters in a more generic situation. Hence, rather than use a particular value, we parametrize our ignorance by allowing it to vary over the entire physical range.

For 300(3000) fb$^{-1}$ luminosity, the discovery reach for {\it Model-$3^F_1$} ranges from 985 GeV to 1140 GeV (1345 GeV to 1485 GeV) with the search sensitivity increasing as the branching fraction of the singly-charged exotics into the neutrino modes decreases. For {\it Model-$5^F_3$}, the discovery reach ranges from 1625 GeV to 1650 GeV (1985 GeV to 2020 GeV) for 300(3000) fb$^{-1}$ luminosity. This substantiates that a significant mass range of the exotics in these simplified models can be probed for discovery at the 13 TeV LHC. Note that, for some of the simplified models such as the {\it Model-$4^F_{1/2}$}, the discovery reach depends moderately on {\it BR}($\chi^\pm \to \nu W^\pm$), while the reach is almost independent for some other models like {\it Model-$5^F_3$}. This is because, for the former, the associated production of the doubly- and singly-charged exotics contributes significantly to the signal, while for the latter, the signal is dominated by the effective pair production of the doubly-charged exotics.

\section{\label{sec:conclusion}Summary and outlook}
The presence of several charged exotic leptons within the new weak gauge fermionic multiplets is a common yet salient feature of many new physics models \cite{Bonnet:2009ej,Liao:2010ku,Bonnet:2012kz,Kumericki:2012bf,Cepedello:2017lyo,Anamiati:2018cuq,Arbelaez:2019cmj,Picek:2009is,Kumericki:2011hf,Kumericki:2012bh,Yu:2015pwa,McDonald:2013hsa,Delgado:2011iz,Ma:2013tda,Ma:2014zda,Ko:2015uma,Avnish:2020rhx,Agarwalla:2018xpc,KumarAgarwalla:2018nrn,Kumar:2019tat,Ashanujjaman:2020tuv,Babu:2009aq,Li:2009mw,Ashanujjaman:2021jhi,Kumar:2021umc} which are well-placed to address various shortcomings of the SM (see Table \ref{table:refofmodels} for an incomplete list of such models and their motivations). Aplenty production of these exotic leptons and their decays to the SM particles offers an up-and-coming way to probe them at the LHC. Owing to their large masses, their eventual decay products---SM leptons and bosons---tend to be highly boosted, with the jets stemming from the SM bosons more likely to manifest themselves as a single fat-jet rather than two resolved ones. Not only are signatures comprising two or three leptons and one or two fat-jets cleaner than the usual LHC searches, but most of these final states also allow for kinematic reconstructions of the exotic leptons. Consequently, such signatures are expected to be sensitive in probing exotic leptons with large masses. In this work, we propose and investigate an appropriate search strategy. Adopting the bottom-up approach for new physics, we consider several simplified models by extending the SM with only one type of new leptonic weak gauge multiplet (triplet, quadruplet or quintuplet) at a time bearing in mind that such simplified models typically arise as low-energy limits of more ambitious scenarios addressing various lacunae of the SM when some of the fields decouple and/or some of the interactions cease to exist. We have performed a systematic and comprehensive collider study for nine such scenarios and estimated the $5\sigma$ discovery reach for each of these for both 300 and 3000 fb$^{-1}$ luminosity of proton-proton collision data at the 13 TeV LHC. The $5\sigma$ discovery reach extends from 985 GeV to 1650 GeV (1345 GeV to 2020 GeV) for 300 (3000) fb$^{-1}$ luminosity for different simplified models. The analysis presented in this work appears to carry through for a large class of simplified models with exotic charged leptons within a new $SU(2)_L$ multiplet. Thus, collider searches, such as the one presented in this work, are anticipated to play a notable role in constraining or testing the new physics scenarios.

\appendix
\section{\label{app:A} Renormalization Group flows in the different models}
In all the simplified models considered in this work, the SM field content is supplemented by new leptonic weak gauge multiplets. The introduction of such multiplets changes the running of the gauge couplings of the $U(1)_Y$ and $SU(2)_L$ groups. At the 1-loop order, the renormalization group equations (RGEs) for the gauge couplings $g_1,g_2$ and $g_3$ of the $U(1)_Y, SU(2)_L$ and $SU(3)_C$ groups can be written as:
\[
\frac{dg_i}{dt} = b_i \frac{g_i^3}{(4\pi)^2}
\]
where $t=\ln \mu$ with $\mu$ being the renormalization point. Within the SM, the coefficients are $(b_1,b_2,b_3) = \left(\frac{41}{10},-\frac{19}{6},-7\right)$. Table~\ref{table:rge} shows the corresponding coefficients in the respective models as well as the energy scales at which the gauge couplings blow up. The existence of a Landau pole below the Planck scale signals that the theory under consideration cannot be an UV-complete one, and must be subsumed in a larger theory that ameliorates the growth of the couplings. Note that Table~\ref{table:rge} includes the effect of only a single pair of the vector-like multiplet (a single chiral field in case of $5^F_0$) and that the inclusion of multiple copies would only bring the scale down. While the inclusion of higher-order terms in the $\beta$-functions would change the location of the Landau pole to an extent, the difference is only quantitative and not qualitative.

\begin{table}[h]
\centering
\scalebox{0.82}{
\begin{tabular}{|l|l|l|l|}
\hline \hline
{\bf Model} & $(b_1,b_2)$ & $\mu_1$ (GeV) & $\mu_2$ (GeV) \\
\hline \hline
{\it Model-$3^F_1$} & $\left(\frac{13}{2},\frac{-1}{2}\right)$ & $6.9 \times 10^{26}$ & -- \\[1ex]
\hline
{\it Model-$3^F_2$} & $\left(\frac{137}{10},\frac{-1}{2}\right)$ & $8.1 \times 10^{13}$ & -- \\[1ex]
\hline
{\it Model-$4^F_{1/2}$} & $\left(\frac{49}{10},\frac{7}{2}\right)$ & $7.5 \times 10^{34}$ & $3.5 \times 10^{25}$ \\[1ex]
\hline
{\it Model-$4^F_{3/2}$} & $\left(\frac{113}{10},\frac{7}{2}\right)$ & $2.5 \times 10^{16}$ & $3.5 \times 10^{25}$ \\[1ex]
\hline
{\it Model-$4^F_{5/2}$} & $\left(\frac{241}{10},\frac{7}{2}\right)$ & $7.5 \times 10^{8}$ & $3.5 \times 10^{25}$ \\[1ex]
\hline
{\it Model-$5^F_0$} & $\left(\frac{41}{10},\frac{7}{2}\right)$ & $1.7 \times 10^{41}$ & $3.5 \times 10^{25}$ \\[1ex]
\hline
{\it Model-$5^F_1$} & $\left(\frac{81}{10},\frac{61}{6}\right)$ & $9.6 \times 10^{21}$ & $1.8 \times 10^{10}$ \\[1ex]
\hline
{\it Model-$5^F_2$} & $\left(\frac{201}{10},\frac{61}{6}\right)$ & $1.6 \times 10^{10}$ & $1.8 \times 10^{10}$ \\[1ex]
\hline
{\it Model-$5^F_3$} & $\left(\frac{401}{10},\frac{61}{6}\right)$ & $1.7 \times 10^{6}$ & $1.8 \times 10^{10}$ \\[1ex]
\hline \hline
\end{tabular}}
\caption{\label{table:rge} The coefficients in the $\beta$-functions and the energy scales $\mu_1$ and $\mu_2$ at which the Landau poles appear in, respectively, $g_1$ and $g_2$ for the different models. For the onset of the RG running, we take $g_1=0.3587$ and $g_2=0.6483$ at $\mu=m_t$, where $m_t=173.35$ GeV \cite{Buttazzo:2013uya}.}
\end{table}

\acknowledgments KG acknowledges support from the DST INSPIRE Research Grant [DST/INSPIRE/04/2014/002158] and SERB Core Research Grant [CRG/2019/006831]. The simulations were supported in part by the SAMKHYA: High Performance Computing Facility provided by Institute of Physics, Bhubaneswar.

\end{document}